\documentclass[12pt]{iopart}

\usepackage{bbm}
\usepackage{txfonts}
\usepackage{soul}

\begin{document}

\title[]{Modeling of experimentally observed topological defects inside bulk polycrystals}

\author{Siddharth Singh$^a$, He Liu$^b$, Rajat Arora$^c$, Robert M. Suter$^{b}$, Amit Acharya$^d$}
\address{$^a$ Department of Civil and Environmental Engineering, Carnegie Mellon University, 5000 Forbes Avenue, Pittsburgh, PA 15213 USA}
\address{$^b$ Department of Physics, Carnegie Mellon University, 5000 Forbes Avenue, Pittsburgh, PA 15213 USA}
\address{$^c$Advanced Micro Devices, Inc. (AMD), 7171 Southwest Pkwy, Austin, TX 78735 USA}
\address{$^d$Department of Civil and Environmental Engineering and Center for Nonlinear Analysis, Carnegie Mellon University, 5000 Forbes Avenue, Pittsburgh, PA 15213 USA}
\eads{\mailto{ssingh3@andrew.cmu.edu}, \mailto{suter@andrew.cmu.edu}, \mailto{acharyaamit@cmu.edu}}
\vspace{10pt}

\begin{abstract}

A rigorous methodology is developed for computing elastic fields generated by experimentally observed defect structures within grains in a polycrystal that has undergone tensile extension. An example application is made using a near-field High Energy X-ray Diffraction Microscope measurement of a zirconium sample that underwent $13.6\%$ tensile extension from an initially well-annealed state. (Sub)grain boundary features are identified with apparent disclination line defects in them. The elastic fields of these features identified from the experiment are calculated.
\end{abstract}


\vspace{2pc}
\submitto{\MSMSE}

\section{Introduction}

The recent development of three dimensional x-ray microscopies that use unit cell orientations as the contrast mechanism \cite{Poulsen2004,Ludwig2009,King2010,Suter2006,Li2013,Bernier2020} has opened up the possibility of making rigorous comparisons between observed and modeled features and responses in interior volumes of metal and ceramic polycrystalline microstructures. For example, comparisons have been made between observed plastic behaviors and model expectations including twinning \cite{Lind2014}, grain rotation and breakup \cite{Pokharel2014,Pokharel2015} as well as crack nucleation and propagation \cite{Gustafson2020,King2008,Hanson2018}. Studies of thermal responses include recrystallization \cite{Liu2022,Hefferan2012,Lauridsen2003}  coarsening \cite{Bhattacharya2021,Bhattacharya2019,Shen2019a,Zhang2020} and phase transformations leading to abnormal grain growth \cite{Higgins2021}. These studies provide direct experimental observations that can and are being used to develop improved models of materials responses. For example, experimental evidence shows that long held assumptions with respect to coarsening are inconsistent with observations based on x-ray microscopy \cite{Bhattacharya2021,Zhang2020}.

Here, a demonstration is presented in which an experimentally observed topologically interesting feature is observed in the orientation field in a near-field high energy diffraction microscopy (nf-HEDM) measurement in the interior of a crystalline grain of zirconium after tensile deformation. The elastic distortion field around this feature is modeled (as a two dimensional system) to show the possible implication of an underlying discontinuity in the orientation field on the stress fields generated in a purely elastic medium. To our knowledge, there do not exist research efforts linking experimental measurements of orientation distributions to the possible stress fields associated with them. 

 Restricting the discussion to the linear theory of elasticity, it is a fundamental result that Nye tensor fields arising from skew symmetric tensor fields cannot produce stress. However, terminating rotation discontinuities or discontinuities with varying magnitude of rotation jump across them can produce stress corresponding to distributions of disclinations and disclination dipoles in such ``interfaces.'' The dipoles are topologically equivalent to interfacial dislocations. We leverage this idea to identify, at the scale of the observations, plausible intra-granular orientation discontinuities and use g.disclination theory and associated computational techniques in two dimensions to compute their stress fields \cite{ZA, ZAP}. Related work can be found in \cite{Sun_fressengeas,sun_fressengeas_2} focusing on defect kinematics where finite difference approximations are used for inferring the disclination density from observations, but no stress fields are calculated. A notable recent work based on EBSD measurements, and broadly complementary to our work is \cite{Demouchy2023}; however, the rare topologically interesting features in the zirconium sample would be difficult to locate with a surface based measurement. The work in \cite{Naragani2021} utilizes grain averaged elastic strain measurements to compute intergranular elastic distortion and stress fields in a polycrystal using field dislocation mechanics theory (a subset of our equations), constraining the computed elastic strain field's grain-wise averaged field to equal the measured grain averaged elastic strain field. 

\section{Experimental data} \label{sec:expt}

Near-field High energy X-ray diffraction microscopy (nf-HEDM) is described in detail in \cite{Bernier2020} which also summarizes a variety of applications. nf-HEDM yields three dimensional maps of crystalline grain unit cell orientations inside polycrystals with micron position resolution and $\sim 0.1$ degree orientation precision \cite{Suter2006,Li2013,Menasche2020}. In deformed microstructures, it has been shown to map intra-granular misorientations \cite{Lind2014,Pokharel2015,Hefferan2012,Renversade2016} and, in suitably optimized datasets, intra-granular strain fields \cite{Shen2020}.

Here, we use the nf-HEDM diffraction image dataset that was used in \cite{Lind2014} to study microstructure evolution in a high purity zirconium sample over a series of states with increasing tensile extension from zero (state S0) to 17.1\% (S3) true strain. The focus here is primarily on data in the S2 state with $13.6\%$ strain. Reconstructions of the orientation field use the HEXOMAP package \cite{Liu2022Git} to determine orientations on a $1.48~\mu\mathrm{m} \times 1.48~\mu\mathrm{m}$ grid of points for each of the 50 measured,  $4~\mu\mathrm{m}$ separated, cross-sections through the $\approx 1~\mathrm{mm} \times 1~\mathrm{mm}$ sample. The tensile extension direction is perpendicular to the reconstructed cross-sections. The undeformed, well-annealed S0 state is also reconstructed in the same way for comparison. The reconstruction code returns a voxel-by-voxel orientation map and an associated confidence parameter, $\mathcal{C}$, for each voxel. $\mathcal{C}$ is defined as the fraction of simulated Bragg peaks that overlap observed intensity in the experimental dataset \cite{Li2013}.  In the elastically and plastically deformed S2 dataset, heterogeneously distributed reduced values of $\mathcal{C}$ are reported due to the use of unstrained lattice parameters in the simulation and heterogeneous broadening of diffraction signals \cite{Li2013,Lind2014}.

\begin{figure}[ht!]
    \centering
    (a)\includegraphics[scale=.48]{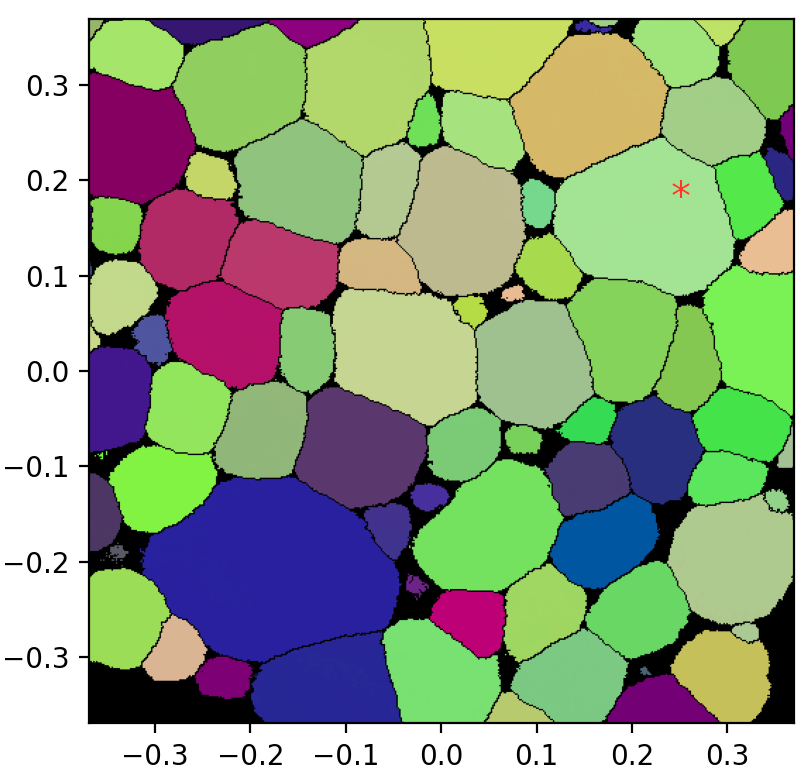} %
    \qquad
    (b)\includegraphics[scale=.48]{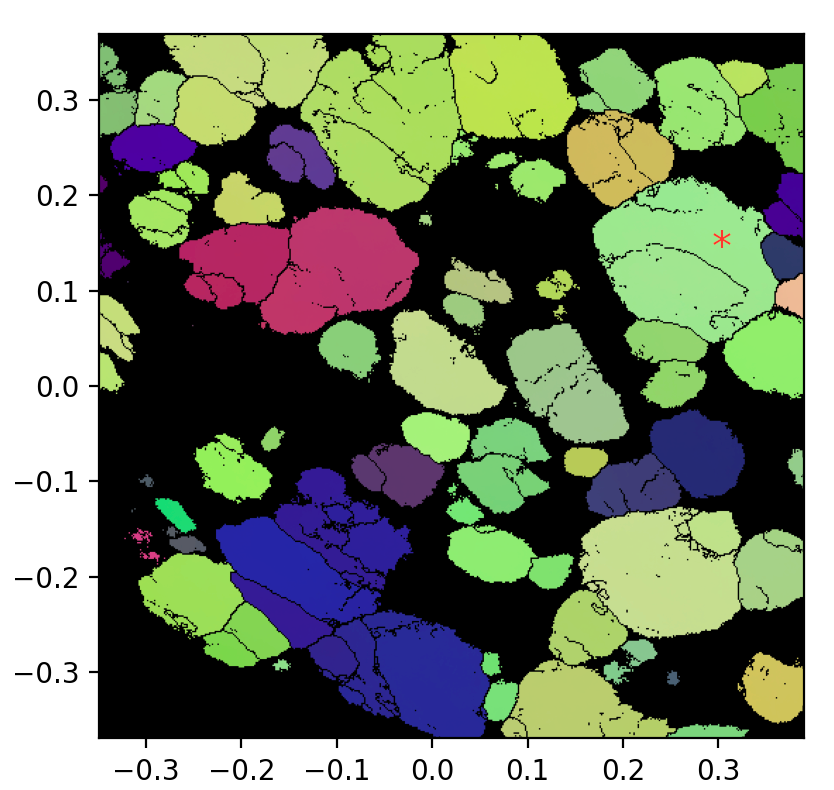}%
    \caption{Corresponding cross-sections through the microstructure in (a) S0 and (b) S2 states. Axis units are in millimeters and colors are mapped from the three parameters specifying unit cell orientations (Rodrigues vector components) The grain of interest is marked with a red '*' in both states. The color scales in these images are identical.
    \label{fig:S0S2}}
\end{figure}

Figure \ref{fig:S0S2} shows matched reconstructed cross-sections from states S0 and S2. The matching process searches the two data sets for measured layers with minimal orientation differences across the data slices; of course, perfect matches do not occur due to the extension and deformation imposed prior to the S2 measurement. As seen in the figure, in S0 the grains have at most very limited internal structure: virtually all reconstructed volume elements (voxels) have orientations within grains that are within the orientation resolution of 0.1 degree. The layer shown in Fig. \ref{fig:S0S2}b was chosen through a search of the three dimensional reconstructed S2 dataset for intra-granular orientation features in which small angle rotations (small angle or subgrain boundaries) occur along a line, presumably due to slip, but in which discontinuities in the rotation values along the line occur within the grain cross-section rather than terminating at a large angle grain boundary.  The grain marked with `*' shows an example; such features are rare in the three dimensional dataset. Fields associated with this topological feature are the subject of the analysis below. Fig. \ref{fig:S0S2}(a) shows the best matched layer to that in (b) in the fully ordered S0 state; this makes it clear that the features seen in (b) are generated by the imposed tensile deformation of the zirconium sample.

In Fig. \ref{fig:S0S2}, only voxels with $\mathcal{C} > 0.4$ are plotted. In (a), the great majority of voxels are shown in orientation coded colors while in (b), extended low confidence areas having $\mathcal{C} < 0.4$ (black) indicate severe deformation and/or substantial elastic strain. In both images, black lines are drawn between all neighboring reconstructed points with $\mathcal{C} > 0.4$ that have orientations that differ by a rotation angle $\Delta > \degrees{1.1}$. In (a), only well defined grains are delineated again indicating well defined crystalline grains. In (b), grain boundaries are still observed and grains that appear in (a) remain, albeit with deformed shapes, but a substantial population of intra-granular misorientations with discontinuities $\Delta > \degrees{1.1}$ also appear.

\section{Modeling of the observed orientation field}

\begin{figure}[ht!]
    \centering
    (a) \includegraphics[scale=.45]{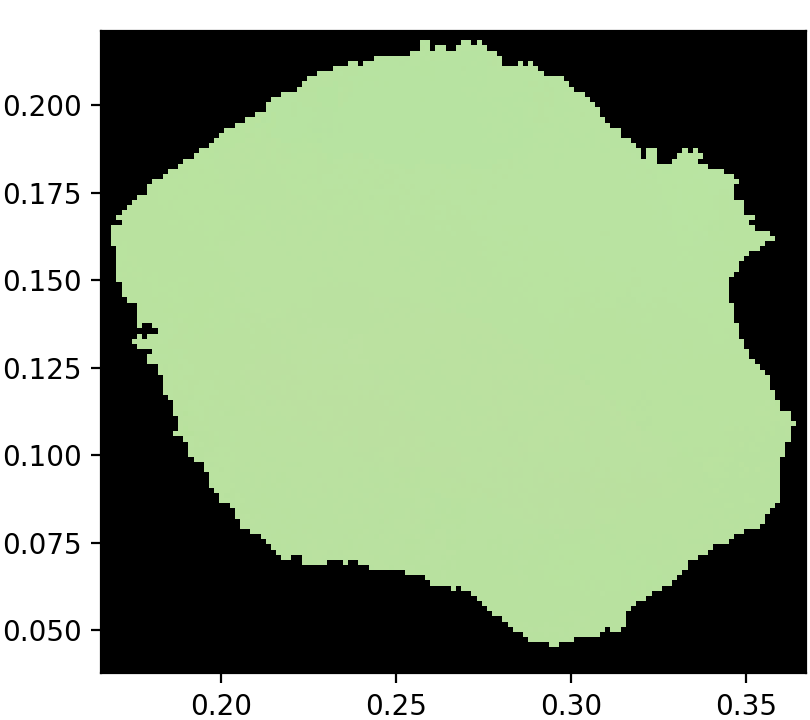}
    (b) \includegraphics[scale=.45]{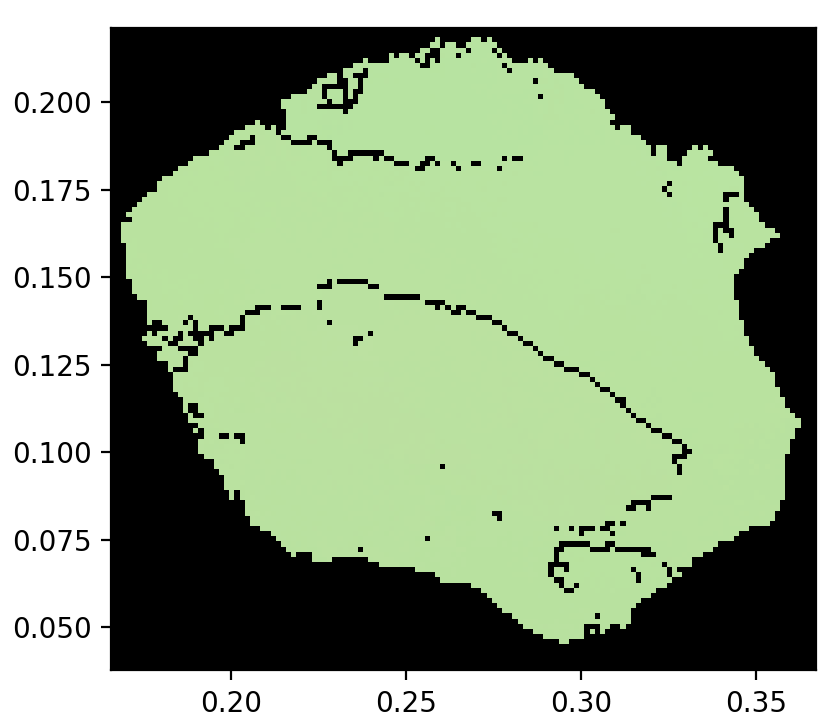}\\
    (c) \includegraphics[scale=.45]{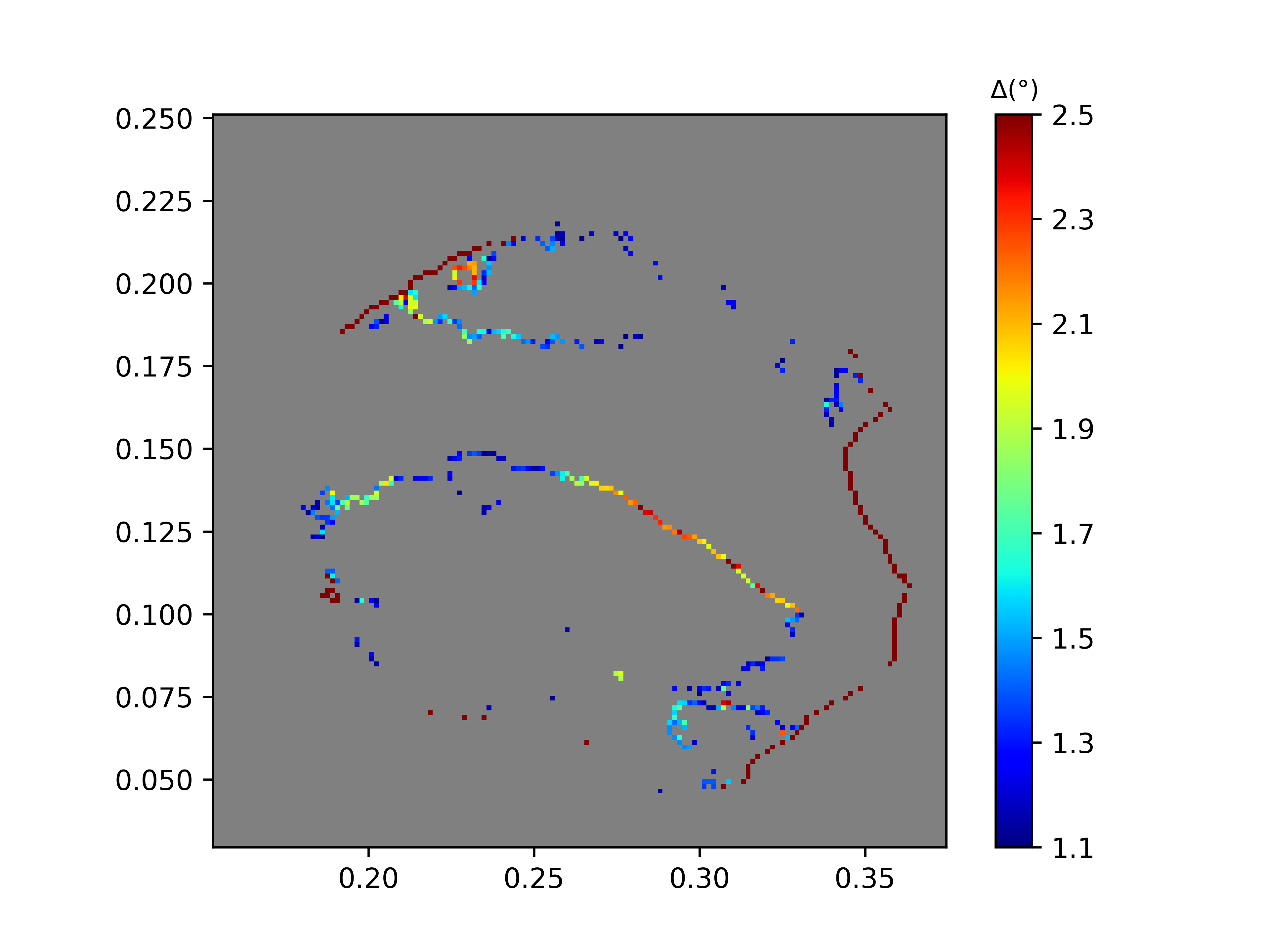}
\caption{The extracted grain cross-section chosen for analysis in state S2. As in Fig.~\ref{fig:S0S2}, black boundary lines are drawn for (a) $\Delta > \degrees{5}$ and (b) $\Delta > \degrees{1.1}$. (c) Shows the value of $\Delta$ along the feature of interest. The orientation of the rotation axis along this line remains almost constant. The small orientation based color variations in (a) and (b) are difficult to distinguish by eye. Note: The background region in (c) has $\Delta < \degrees{1.1}$ }
\label{fig:thegrain}
\end{figure}

The grain of interest seen in Fig.~\ref{fig:S0S2} is shown as extracted from its surrounding grains in state S2 in Fig.~\ref{fig:thegrain} which also demonstrates that the orientation field in this grain even in S2 is relatively uniform, having no discontinuities larger than $\degrees{5}$ (panel (a)).  But, well defined discontinuity lines are seen in (b) with the lower threshold of \degrees{1.1}.  This verifies the small angle nature of all interior boundaries and the relative uniformity of orientations around them.

Our goal here is to demonstrate a procedure to estimate the linear elastic stress fields associated with what are variously referred to as low angle, small angle or sub-grain boundaries as they are observed in HEDM data. We denote these boundaries as s-gbs. Such s-gbs are usually thought of as being comprised of a dislocation wall but, noting that dislocations can be decomposed into disclination dipoles, we adopt the language of disclinations. Individual disclinations store high levels of elastic energy in the far-field region away from the defect core and therefore usually appear in dipole combinations. However, our calculations are sensitive to longitudinal variations along boundaries of the rotation jump across the boundary. An extreme example of such a variation is if the boundary terminates within a grain; a termination corresponds topologically to a bare disclination. Similarly, less extreme variations observed along the boundary (Fig. \ref{fig:thegrain}(c)) correspond to a continuous distribution of disclination density.

All the defects discussed above occur on the nano-scale and are not explicitly observed in the nf-HEDM measurements. However, the longer length scale consequences of collections of defects yield the unit cell rotations that are observed at experimental resolution. We surmise, as is commonly done, that nano-scale defect arrays underlie the s-gbs. The computations yield stress fields on the meso-scale length scales of the measurements.

In this section, we describe our interpretation of the experimental data, introduce the computational model, and describe the stress fields obtained from the computations.

\subsection{Interpretation of voxelized experimental data}\label{sec:Smooth_sgb}

\begin{figure}[ht!]
    \centering%
    (a) \includegraphics[scale=.55]{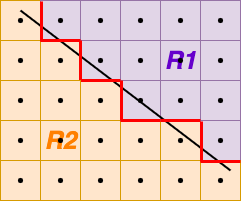}%
    \hspace{1cm}
    {(b) \includegraphics[scale=.55]{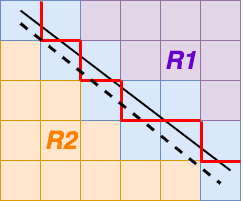}}%
    \caption{ Outline of s-gb identification for computations: (a) Black dots are points at which the reconstruction software determines rotation values (lattice orientations). The analysis associates rotation values with the area of the corresponding square box. Two different rotations are shown on either side of an actual s-gb (bold black line). The stepped red line represents the s-gb as identified by the measurement. (b) For modeling, we identify boxes (marked as light blue) across which there exist large rotation gradients.
    We obtain the s-gb shape by a smooth curve (dashed black line) passing through the voxel centers.
z    }%
        \label{fig:sgb_prof}%
\end{figure}

With reference to Fig.~\ref{fig:sgb_prof}, the nf-HEDM data identifies an interface as the jagged curve (colored in red) formed by the boundaries of the voxels across which an above-threshold orientation discontinuity exists. In the modeling, we represent the interface by a smooth curve passing through the voxels, abutting the jagged curve, which correspond to regions of high gradients in measured orientations (described by the magnitude of the field $P$ subsequently defined in (\ref{eq:P})). The smoothed interface so defined is assumed to be the surface/curve across which the orientation jumps. For the s-gb considered in this study, this smooth curve is obtained by a $5^{th}$ order polynomial fit of the relevant set of voxel centers defining the s-gb.

In this interpretation of the experimental data, no orientation (or distortion) can be assigned to the microscopic interface itself (the red jagged curve) and therefore the tangential gradient of orientation/distortion along its smooth representation is assumed to vanish; this assumption has no influence on our results. Thus, the only measurement attributable to the interface is an assertion of the normal gradient of orientation across it, given by the jump in orientation/distortion, divided by the width of the boundary, an input to the theory. This normal gradient field constitutes the eigenwall field, $S$, of the theory defined in Sec.~\ref{sec:g.disc_theory}. The theory suggests that any region of the interface where the orientation/distortion jump varies longitudinally (i.~e., along the dashed line in Fig.~\ref{fig:sgb_prof}(b)) serves as a source for stress (a location of non-zero $\mathit{\Pi}$ field). For example, in a twin boundary with curvature, this must be true as the distortion jump depends on the normal. We note that a tangential variation, of the jump in orientation normal to the interface, is not related in any way to assigning orientations to the interface, or a tangential gradient of the orientation field on the interface. In our calculations, we will be conservative and assign the width of the boundary to be a voxel length, having the effect of predicting lower stresses due to the identified defects. The calculations in this paper involve linear partial differential equations (pde), so this amounts to a scaling of stress fields by a constant factor.

\subsection{Elements of g.disclination theory}\label{sec:g.disc_theory}
We review here the static elastic g.disclination theory of \cite{ZA, ZAP}. The (inverse) elastic distortion and the eigenwall fields are the main kinematic ingredients of the model. A disclination is defined as the terminating curve of an interface supporting a discontinuity in the elastic rotation. A g-disclination is more general, and refers to the terminating curve of an interface supporting a discontinuity in the (inverse) elastic distortion. Unlike interfaces of (inverse) displacement discontinuity which may or may not be physically identifiable (e.g., identifiable stacking faults, or `perfect' 1-atom spacing shifts across an atomic layer otherwise), (inverse) elastic distortion discontinuities (at a macroscopic scale) are always identifiable. If $W$ denotes the (inverse) elastic distortion, at an interface like a phase or grain boundary it has a large gradient over a thin interfacial region; we think of this concentrated gradient as the `singular part' of $W$ and refer to it as the \textit{eigenwall} field $-S$. The singular part is separated out of the (distributional) derivative of $W$, $DW$, to define the `regular part' of $DW$ as
\begin{equation}\label{eq:Y}
    Y := DW - (-S) = DW + S.
\end{equation}
$Y, DW, S$ are third order tensor fields.

When the elastic distortion $U^e$ is small, $W \approx I - U^e$, in which case (\ref{eq:Y}) can be expressed as
\[
Y = - DU^e + S.
\]
With $X$ as the alternating tensor and the \textit{definition} $Y:X =: \alpha$,
\begin{equation}\label{eq:alpha}
Y:X - S:X = \mbox{curl} \, U^e \qquad \Longrightarrow \qquad \alpha + (-S:X) = \mbox{curl}\, U^e.
\end{equation}
Since both terms on the l.h.s.~of (\ref{eq:alpha}) are sources of incompatibility of the elastic distortion  (since $\mbox{curl}\,U^e = 0$ implies that $U^e$ is a gradient in simply connected domains),  (\ref{eq:alpha}), along with the physical interpretation of the eigenwall field, says that $-S:X$ is the \emph{interfacial dislocation density} and $\alpha$ is the \emph{bulk dislocation density} field. \textit{Both serve as sources for elastic distortion} of the body causing stress (in many, but not all, circumstances).

Another important source of elastic distortion is the \emph{g.disclination density} $\mathit{\Pi}$ defined as
\[
\mbox{curl}\, Y = \mbox{curl}\, S =: \mathit{\Pi}.
\]

In the following we define $\mathit{\Omega}$ as the domain of interest in which we want to calculate elastic distortion fields and stress due to interfacial dislocation density fields (and possibly bulk dislocation density fields). As explained in \cite{ZAP}, in order to deal with stress fields arising from longitudinal gradients of the eigenwall field along grain and compatible phase boundaries, i.e., g.disclination density fields, we define
\begin{eqnarray*}
        \eqalign{\mbox{curl}\, {S}^\perp & = \mbox{curl} \, {S} \qquad \mbox{on} \  \mathit{\Omega} \\
         \mbox{div}\,  {S}^\perp & = 0 \qquad \mbox{on} \  \mathit{\Omega} \\
          {S}^\perp n & = 0  \qquad \mbox{on} \  \partial  \mathit{\Omega} \\
         -\mbox{div} \, D  {H} & = \mbox{div}\,  {S} \qquad \mbox{on} \   \mathit{\Omega} \\
         (-D  {H}) n & =  {S} n \qquad \mbox{on} \  \partial  \mathit{\Omega}. }
\end{eqnarray*}

so that
\[
{S} =  {S}^\perp - D  {H}.
\]

Then, with the definition $ \hat{U}^e := U^e +  {H}$, (\ref{eq:Y}) may be equivalently stated as
\begin{equation}
    \label{eq:Y_rew}
    Y - {S}^\perp = - D \hat{U}^e
\end{equation}
so that
\begin{equation*}
    \alpha -  {S}^\perp:X  = \mbox{curl} \left(U^e +  {H} \right) =: \mbox{curl}\, \hat{U}^e.
\end{equation*}
Finally, defining the constitutive equation for stress to be $\sigma = \mathbbm{C} \,  \hat{U}^e$, where $\mathbbm{C}$ is the tensor of (isotropic or anisotropic) linear elastic moduli, the elastostatic field equations at `small distortions' become
\begin{eqnarray}
    \label{eq:GT_feq}
    \eqalign{\alpha - {S}^\perp:X & = \mbox{curl}\, \hat{U}^e\\
     \mbox{div} (\mathbbm{C}\, \hat{U}^e) & = 0 \qquad \mbox{on} \  \mathit{\Omega}\\
    (\mathbbm{C}\, \hat{U}^e)n & = t \qquad \mbox{on} \  \partial  \mathit{\Omega},}
\end{eqnarray}
where $t$ is a statically admissible applied traction field (possibly vanishing).

The l.h.s. of (\ref{eq:GT_feq}) shows the sources of elastic distortion. It can be shown \cite[Sec.~6]{ZAP} that when $S$ is skew in its first two indices (as happens with the modeling of low angle s-gbs, as is the case here) the elastic strain and stress depend only on the knowledge of the bulk dislocation density $\alpha$ and the g.disclination density $\mathit{\Pi}$. It can also be inferred that, under the same assumption, if $S^\perp$ is replaced by $S$ and $\hat{U}^e$ by $U^e$ in (\ref{eq:GT_feq}), then $\hat{U}^e_{sym} =  U^e_{sym}$ and the stress and elastic strain remain unchanged.
However, the rotation fields $\hat{U}^e_{skw} \neq U^e_{skw}$, if $S \neq S^\perp$.

\subsection{Computation of elastic fields of grain and sub-grain boundaries}
\label{sec:modeling}

\begin{figure}
    \centering
    (a) \includegraphics[scale=.45]{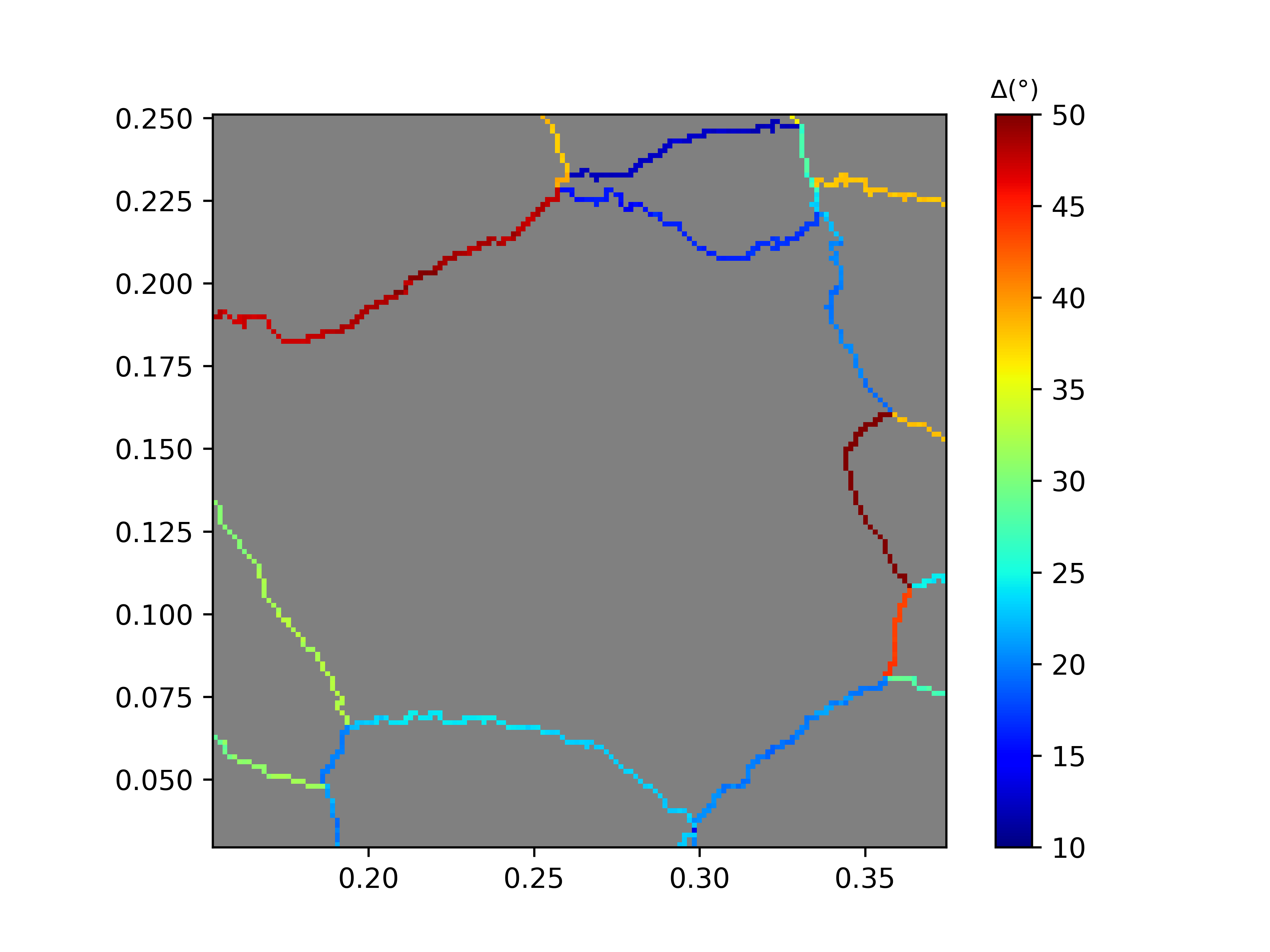}
    (b) \includegraphics[scale=.45]{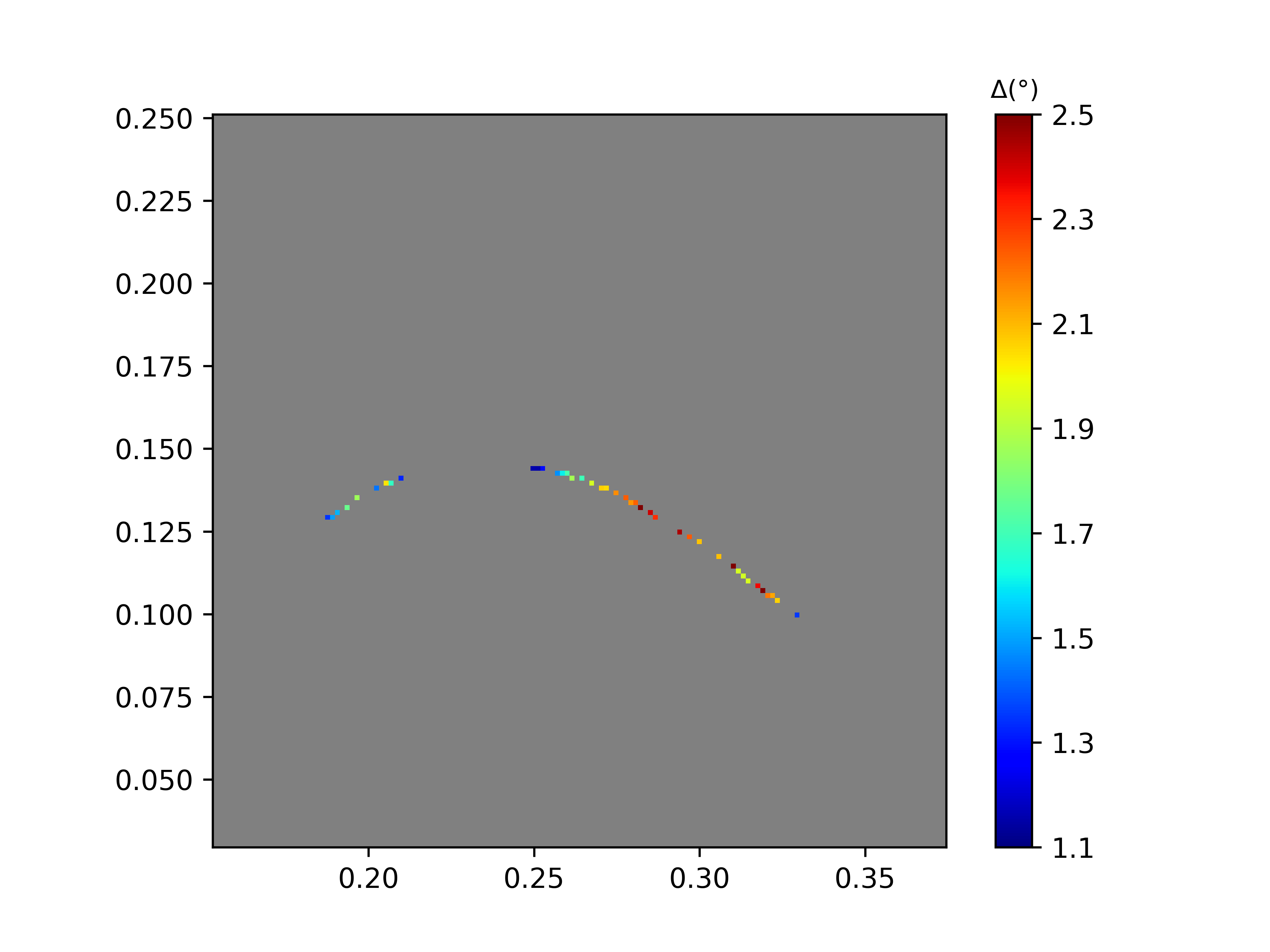}\\
    (c) \includegraphics[scale=.45]{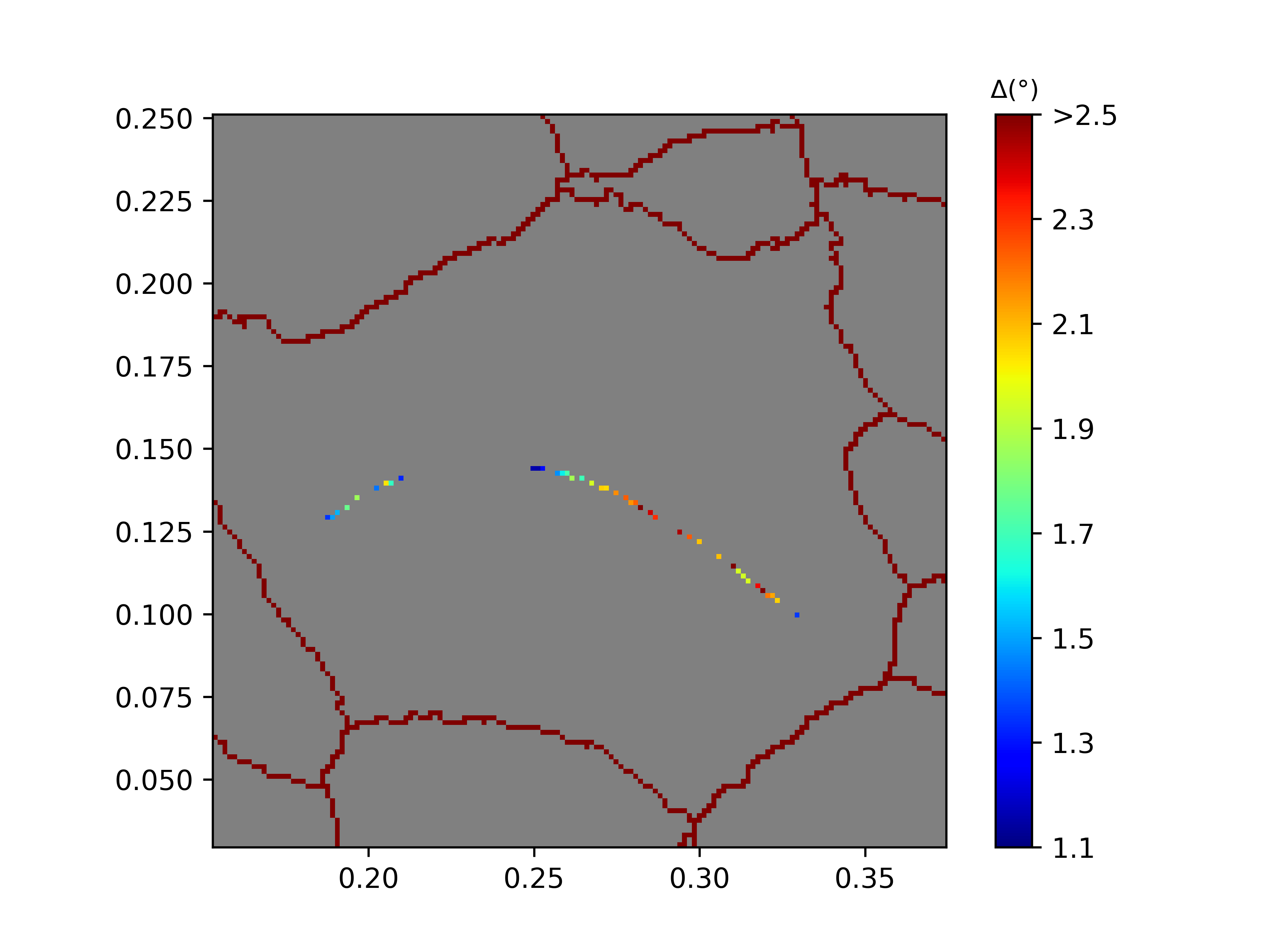}
\caption{Misorientation (in degrees) for (a) the network of grain boundaries, (b) the s-gb and (c) the combined system for which the stress field is calculated. Different colorbar scales are used to make features more evident. The background regions in all the figures have $\Delta < \degrees{1.1}$ }
    \label{fig:modeling}
\end{figure}

In this section, we discuss the approach adopted in this paper to calculate the elastic fields of an interfacial network comprising grain boundaries (high-angle boundaries) and s-gb (low-angle boundaries).  Grain boundaries have a large angle of misorientation in comparison to s-gbs within grains, and form a continuous network without terminations. On the other hand, unlike grain boundaries, s-gbs may be accompanied by high gradients of misorientation in the longitudinal direction along the feature, as evident in Fig.~\ref{fig:modeling}.

\begin{figure}[ht!]
    \centering
    (a) \includegraphics[scale=.45]{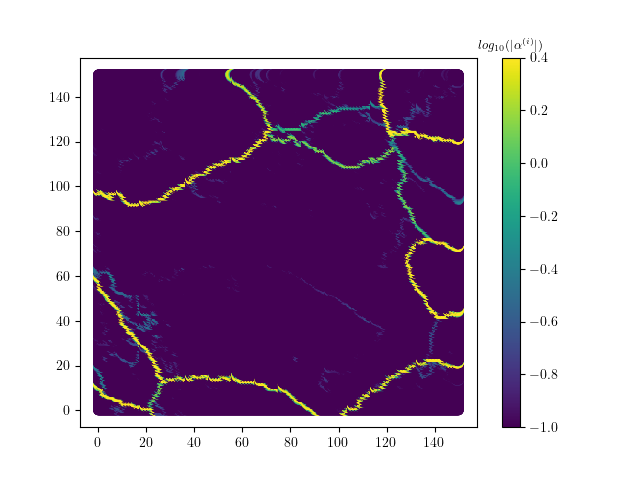}%
    (b) \includegraphics[scale=.45]{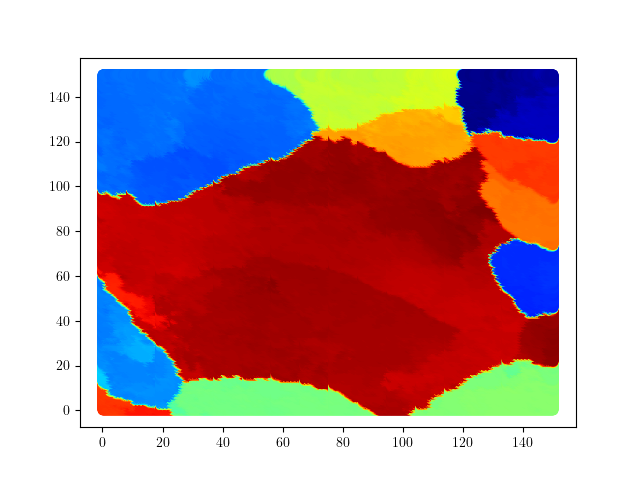}
    \caption{(a) shows the interfacial dislocation density for the network of grain boundaries; $\alpha^{(i)}$ is defined in \ref{app:A}. For measurements with satisfactory certainty, the (terminations in the) s-gb are not captured. 
  (b) is the rotation field considered for the simulations. The grains are colored based on the magnitude of the Rodrigues vector.}
    \label{fig:calalpha}
\end{figure}

Significant longitudinal gradients along a gb can occur only at junctions where another gb meets it, by definition. If these transverse gbs are not taken into account, the large longitudinal gradients along a gb necessarily produce large stresses, as the `sites' of such longitudinal gradients serve as (continuously distributed) disclinations that act as sources of elastic distortion with non-trivial strain. For a realistic prediction of stresses within and around a single grain due to interfacial defect fields, it is thus important to account for the network of grain boundaries connected to a single grain.connected to a single closed gb. Moreover, if only rotation can be measured, and the presence of strain can make the measurement uncertain, then for higher degree of confidence (defined in Sec.~\ref{sec:expt}) the grain boundary network may not appear connected as shown in Fig.~\ref{fig:thegrain}(c). For this practical reason related to modeling with experimental data, we reduce the confidence value of reconstruction used in Fig.~\ref{fig:thegrain}(c) set at 0.4 to 0.2 to obtain the gb network shown in Fig.~\ref{fig:modeling}(a). The corresponding stress field of the gb network is discussed at the end of Sec.~\ref{sec:simu_result}.

The theoretical framework employed in this first paper for computing stress of experimentally measured interfacial features as outlined in Sec.~\ref{sec:g.disc_theory} applies to the small elastic distortion case (the theory and calculations in \cite{ZAP} are not so restricted). However, the presence of a grain boundary, as discussed, implies large misorientations and hence large elastic rotations, and the utilization of the aforementioned linear theory for calculation of stress is not appropriate. As a compromise, instead of using a nonlinear elastic constitutive relation for stress and an interfacial dislocation density field calculated from a finite rotation tensor field with large misorientation representing the gb network, we define an infinitesimal rotation tensor field $ \omega^e := X \Phi$ from the measured finite rotation vector field $\Phi$ (a vector with magnitude representing the angle of rotation in radians and direction giving the axis of rotation) and calculate its stress field; clearly, $\mbox{curl} \, \omega^e$ concentrates on the gb network and it produces small stresses as would the fully nonlinear theory without approximation, while identifying the gb network accurately. As a validation of our approach, we have plotted the $\log_{10}$ of the norm of the interfacial dislocation density (Fig.~\ref{fig:calalpha}a) obtained from this $\omega^e$ field and we note that grain boundaries are observed quite well.

For the s-gb, high gradients in the interfacial dislocation density along the s-gb is observed in the experimental data (Fig.~\ref{fig:thegrain}) and consequently should show up in the modeling. An s-gb, by definition, comprises a `small' misorientation and, with the additional (self-consistent) assumption that the elastic strain field produced by such s-gb features is small (which, nevertheless, can produce stresses that are two to three orders of magnitude higher than typical macroscopic yield stresses), the linear elastic theory discussed in Sec.~\ref{sec:g.disc_theory} can be used to compute its stress field. Then, the overall stress field of the gb network combined with the s-gb is obtained by superposition due to the use of linear elasticity for calculation of stresses.

Mathematically, let $F = R \sqrt{C}$ represent the unique pointwise polar decomposition of the (finite) elastic distortion tensor into an orthogonal elastic rotation tensor $R$ and the right stretch tensor $\sqrt{C}$, and we define $\varepsilon := \sqrt{C} - I$. Then the inverse elastic distortion tensor field is $W = (I + \varepsilon)^{-1} R^T$. In any given situation, the actual strain field $\varepsilon$ is non-vanishing in general, but we assume that only the rotation field $R$ is measured. We now assume that across an (s-)gb with normal $n$ the rotation field can appear discontinuous with values $R_t, R_b$ while the strain field is continuous. Then across such an (s-)gb, the difference in elastic distortion is given by
\[
\Delta W = (I + \varepsilon)^{-1} \left( R_t^T - R_b^T \right) \approx \left( R_t^T - R_b^T \right) - \varepsilon \left( R_t^T - R_b^T \right)
\]
with the approximation valid when $|\varepsilon| \ll 1$. In such a case the source of elastic distortion, \textit{including the elastic strain}, according to g.disclination theory is $S:X$ with $S$ well approximated by 
\[
S = \left( R_t^T - R_b^T \right) \otimes n,
\]
and we have the interesting result that \textit{observations of the (finite) rotation field alone, without measurement of strain, can provide the source for internal stress to a good approximation in many situations of physical relevance}. The magnitude of this source is small for an s-gb,  and hence the use of the linear theory for calculation of stress is justified. For a gb, this magnitude of the eigenwall field is large and the linear theory for computing stress is not justified, and we resort to the device mentioned in the previous paragraphs.

Two remarks are in order here: first, the computational implementation of $\Delta W$, the field $P$ in Sec.~\ref{sec:2-d approx}, succeeds in identifying (s-)gbs with terminations that are in reasonable accord with those in the experimentally observed misorientation field, where the latter does not involve any construct of g.disclination theory, see Figs.~\ref{fig:magP} (a) and ~\ref{fig:thegrain} (c). Second, the result in \ref{app:A} implies that if the linearized formula $X:\Phi$ instead of $\exp(X:\Phi)$ corresponding to a finite rotation vector field $\Phi$ is used to define the rotation field $R$ utilized above, it would not be possible to quantitatively identify terminating interfaces and their stress fields (as measured by the $\mbox{curl} \, \alpha^T$ field). 

\subsection{Definition of the input to the computational model}\label{sec:2-d approx}
In this first demonstration, we restrict ourselves to calculations of elastic fields that vary only in two dimensions and, of course, we choose the measured plane discussed above in Sec.~\ref{sec:expt}. For modeling, the restriction to the plane requires the definition of rotations with axes perpendicular to the chosen plane.  In the following, we describe the procedure for defining this input. We note that the restriction to two dimensions is only for simplicity - our theory and experimental techniques are appropriate for 3-d analysis and finite deformation elastic field analysis.

The theory works with the inverse elastic distortion tensor, $W$, and we assume that discontinuities in inverse elastic distortion occur only through its rotation field. Thus, we focus on the measured orientation field whose transpose gives the rotation of the inverse elastic distortion. 
To convert the $3\times3$ rotation matrix data from the measurement for our 2-d analysis, we retain the upper-left $2\times2$ sub-matrix of the $3\times3$ transpose of the rotation matrix written in the sample basis. The distortion field, $W$, so obtained is not a pure rotation field, but this is not an impediment for our elastic field calculations 
as it can treat full distortion discontinuities and not just rotational ones. 

\begin{figure}[ht!]
    \centering%
    \includegraphics[scale=.48]{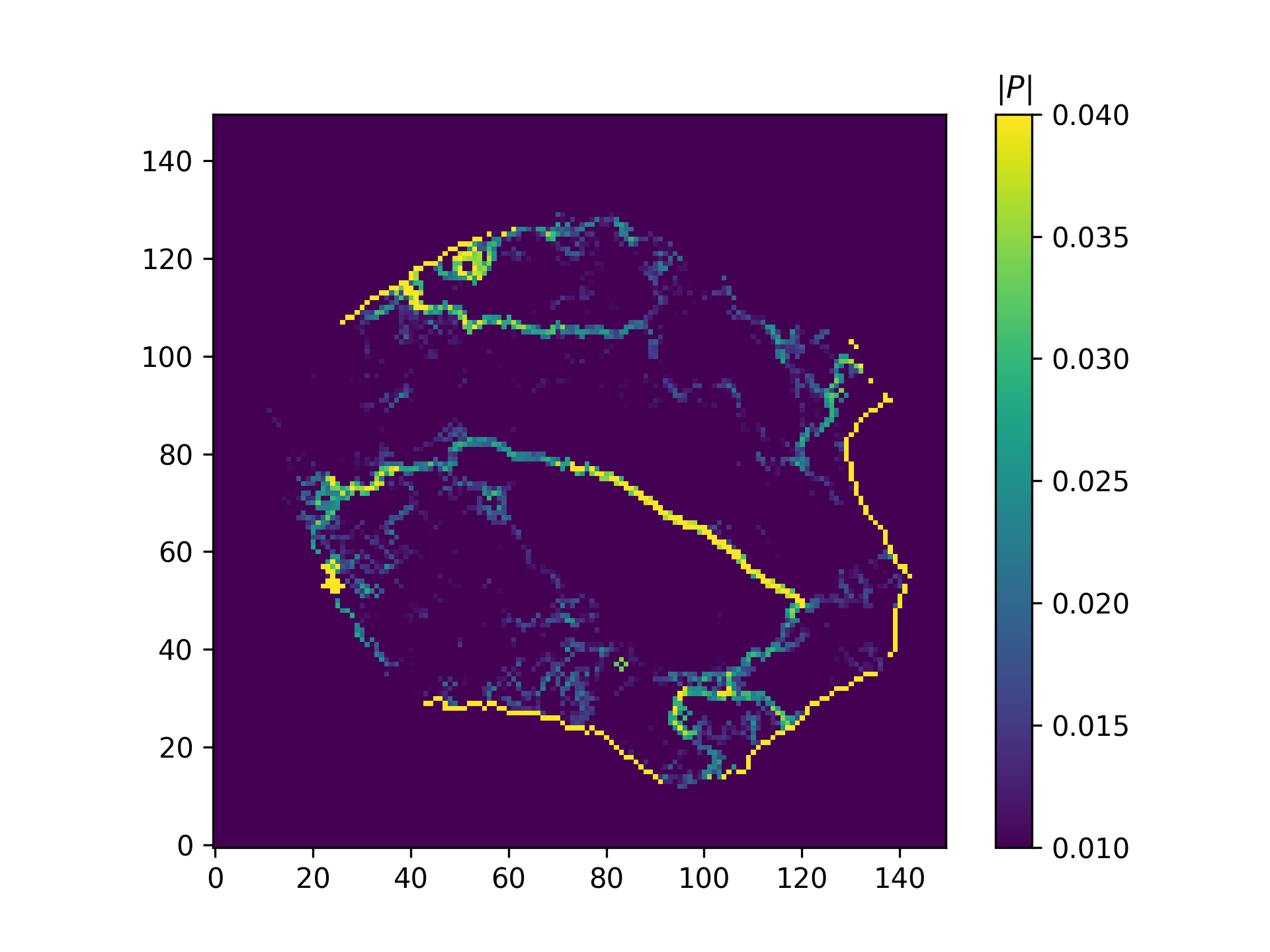}
    \rput(-7.5,5){(a)}
    \includegraphics[scale=.48]{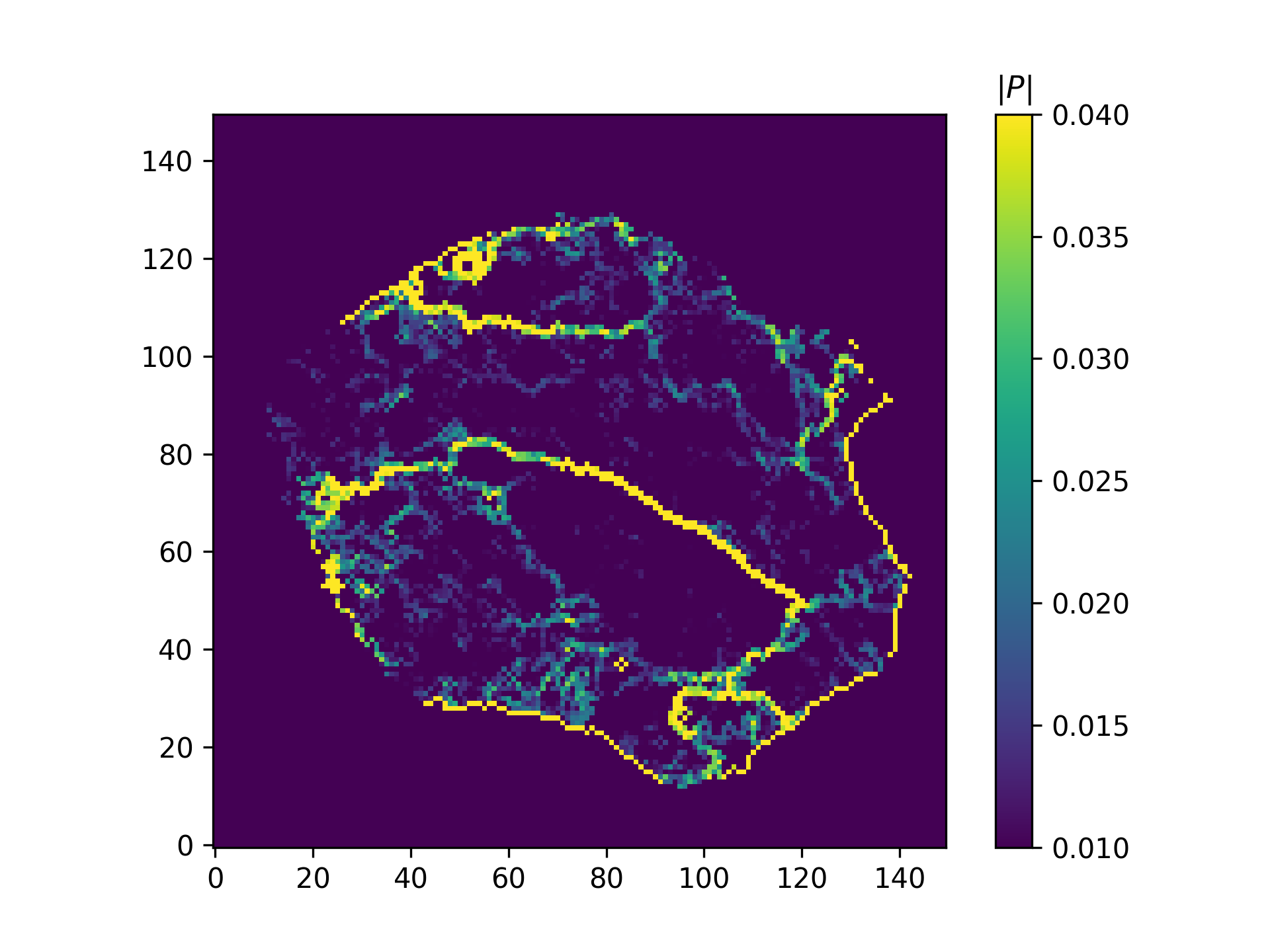}%
    \rput(-7.5,5){(b)}
    \caption{ Color map of $|P|$ for the selected grain of interest. For comparison, (a) shows $|P|$ for the reduced 2-d (inverse) elastic distortion field while (b) shows $|P|$ for the full 3-d rotation field. Scale: 1 unit = 1.48$\um$.}
        \label{fig:magP}%
\end{figure}

With the 2-d distortion field in hand, we define a field $P$ as a discrete gradient of $W$:
\begin{eqnarray}
\label{eq:P}
        \eqalign{DW_i(x) &= W(x + h e_i) - W(x - h e_i)\\
        P &= \sum_{i=1}^{2} DW_i \otimes e_i.}
\end{eqnarray}
$P$ captures the idea of how $W$ varies in all directions. The value of $P$ in the voxel of interest is calculated from the difference in the 2-d distortion field values in the adjacent voxels in two orthogonal directions. $P$ contains information about the discontinuity in $W$ at the voxel level. To characterize the strength of discontinuities, we define the magnitude of $P$ as $|P| = \sqrt{P_{ijk}P_{ijk}}$. Fig.~\ref{fig:magP} shows maps of $|P|$ in the selected grain shown in Fig.~\ref{fig:thegrain}. For comparison, $|P|$ is shown based on both the 2-d and 3-d rotation specifications. In both cases, internal s-gbs appear as expected from Fig.~\ref{fig:thegrain}. Smaller values of $|P|$ also appear frequently as continuous features. 

To define a continuous distortion field at surrounding grain boundaries, distortion data corresponding to voxels having low confidence values ($\mathcal{C}$ in the reconstruction) were included. Low $|P|$ magnitude along some parts of the gb is attributed to the fact that the low $\mathcal{C}$ value orientations in voxels around those parts is nevertheless similar to that inside the grain. As discussed above, these voxels are not included in the grain definition because of their low confidence value. It is also observed that $|P|$ is smaller for the s-gb when compared to gb as is expected. 
Discontinuities that appear in the s-gb in the 2-d case that are less obvious in the 3-d case, are most likely because of changes in the dominant rotation axis. Low-level features are prevalent in 3-d cases because full rotations are not restricted to the single, perpendicular axis.

\subsubsection{Construction of the eigenwall field and g.disclination density from experimental data} \label{subsec:Cons_eig}
\ 
 
\begin{figure}[ht!]
\centering%
    \includegraphics[scale=.48]{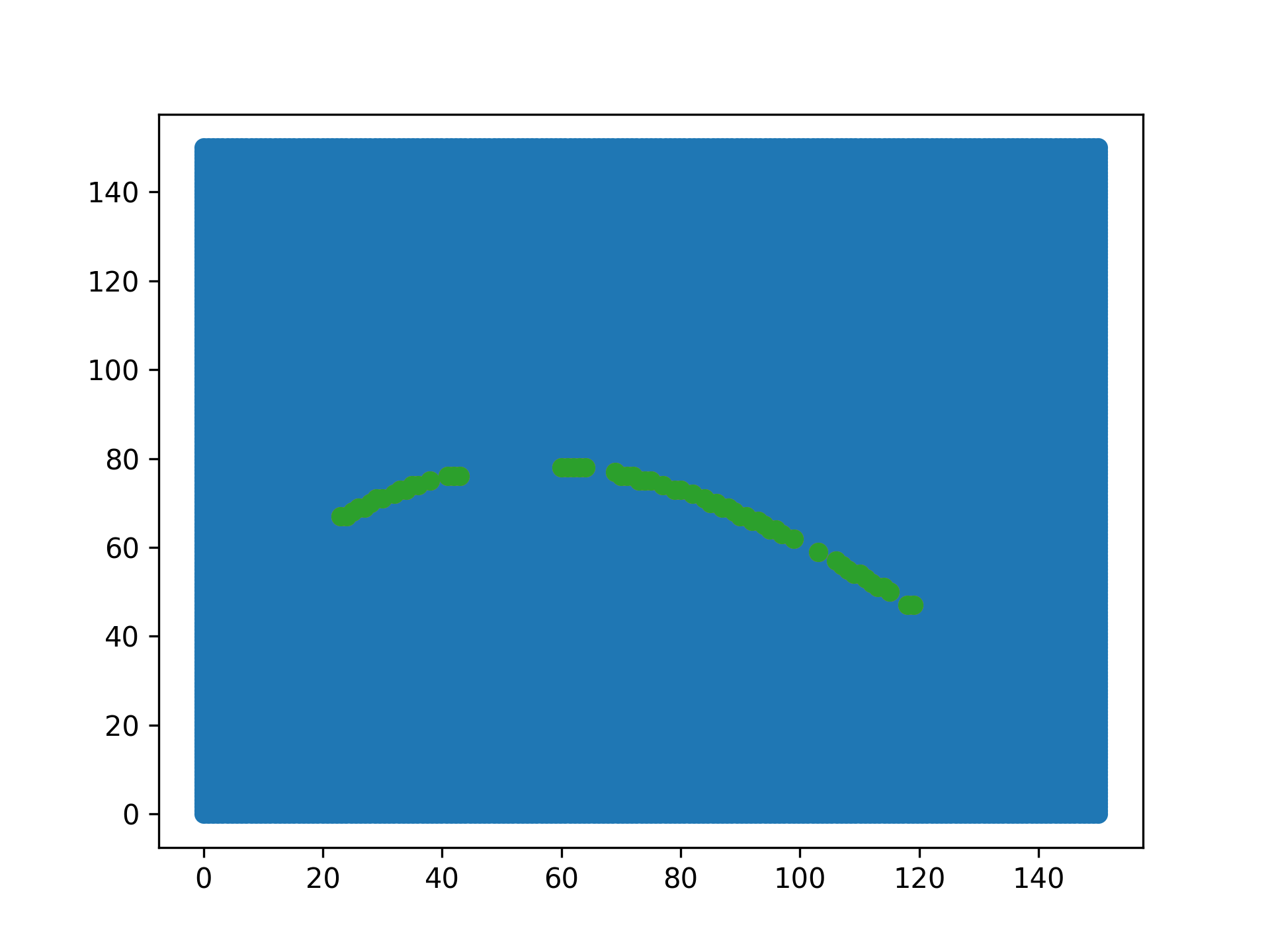}
    \rput(-8,5){(a)}
        \includegraphics[scale=.49]{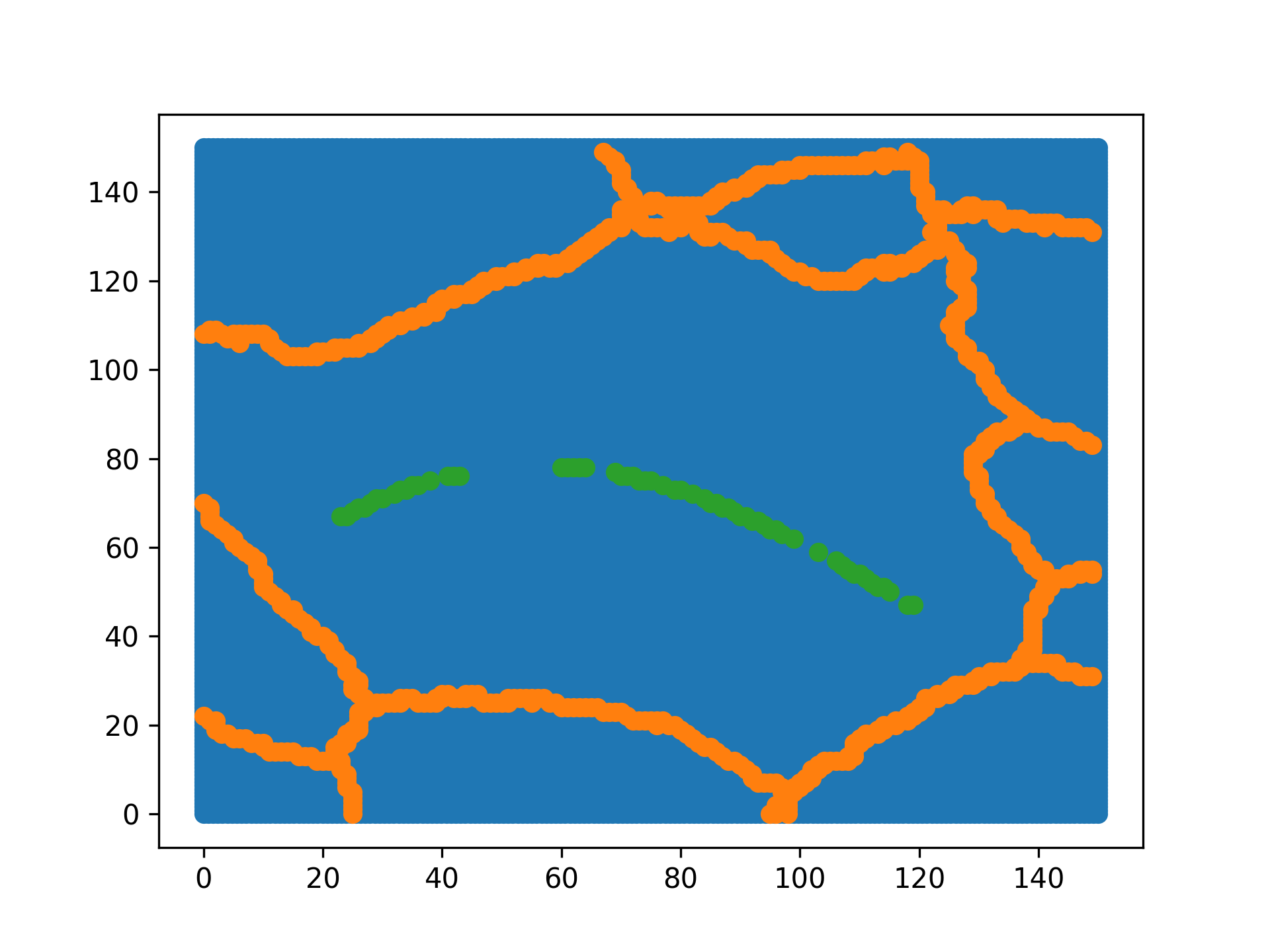}
    \rput(-8,5){(b)}
    \caption{ Simulation domain containing (a) just the s-gb with a prescribed Eigenwall field. (b) s-gb and network of grain boundaries having interfacial dislocation density.
    The elastic computational domain is shown in blue.}%
        \label{fig:padding}%
\end{figure}

With the observation that s-gbs have larger $|P|$ than background features, we set a threshold significantly above the background level and extract voxels with values of  $|P|$ above the threshold. These voxels are then used to define s-gbs. The value of $P$ at other voxels is set to zero, as these values are not considered for any further calculations. A single s-gb feature, shown in Fig.~\ref{fig:padding}, is used to simulate the elastic fields. Given the smoothed s-gb shape, we next calculate the value of the distortion jump, or eigenwall field $S$, to be assigned to each voxel.

The eigenwall field $S$ takes third-order tensors as field values and provides a discrete directional derivative of the orientation, in the direction of the normal, $n$, to the s-gb. For the calculation of elastic fields, we need the eigenwall field $S$ on the Gauss integration points of an FEM mesh. The algorithm for generating such an $S$ field is presented next. The details are presented in the context of a rectangular mesh, but the idea can be extended to general non-uniform meshes.

\begin{itemize}
    \item Voxel centers, where $P$ is evaluated, are considered to be nodes of an FEM mesh. The $S$ field is then evaluated at the Gauss points of this mesh, with the $P$ field as given on the nodes of the mesh.
    
    \item As discussed in Sec.~\ref{sec:Smooth_sgb}, a smooth curve representing the interface (s-gb, gb)  interpolates the centers of the voxels. This, along with the previous point implies that elements in the interior of an s-gb have exactly 2 nodes where $|P|$ is non-zero.
    
    \item To model the termination, we first need to identify the element(s) that form the termination. To do so, we search within the list of nodes which have non-zero $|P|$ values - also terms as interfacial nodes - for one that is not in the interior of the s-gb profile (defined by the criterion above). Such a node is defined here as an end-node. To identify the end-node, we pick a node, say $A$, in the interfacial list and for all elements that have this node on their boundary, we check each for the presence of another node (on their boundaries) with a non-zero $|P|$ value (i.e. belonging to the interfacial list). If this element set, corresponding to $A$, contains more than one element, then this node is considered to be an end-node; otherwise $A$ is deemed to be an interior node. Next, we identify the element(s) that constitute the termination. The idea is to declare that element(s) as the termination, within the element set which shares the end-node as a bounding node, which contains the most aligned line segment to the s-gb profile. As an example, consider Fig.~\ref{fig:term_alg}. Yellow nodes are those in the interior of the s-gb. The purple node represents the end-node, and the green colored elements are the family of elements which can contain the termination. The angle between each dotted line connecting the purple and pink nodes, and the line segment connecting the purple and yellow node on the s-gb is calculated. Whichever dotted line segment has an angle closest to $\degrees{180}$ is deemed as the element containing termination. It is important to note that two dotted line segments might correspond to the same angle. In that case, we consider both elements as constituting the termination.

    \begin{figure}[ht!]
\centering
    \includegraphics[scale=.70]{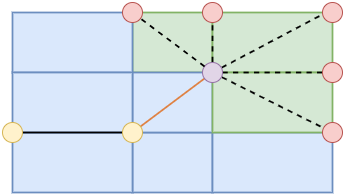}
    \caption{Schematic explaining the algorithm for modeling terminations: Yellow nodes depict the s-gb interior, purple nodes mark the "end node," red nodes indicate potential termination nodes. Green elements form a group that includes termination. Orange and solid black lines outline the s-gb on the mesh.}
    \label{fig:term_alg}%
\end{figure}

    \item The normal to the s-gb, $n$, is evaluated locally by performing a cross product of the tangent vector along the s-gb (formed by joining the two nodes of the element along the s-gb) with the unit vector perpendicular to the plane and then normalizing it to make it a unit vector.
    \item We interpolate the value of the $P$ field to the Gauss points using standard FEM shape functions. If the s-gb lies on a common edge of two elements, the Gauss point values of $P$ in the elements sharing this edge are reduced to half their original interpolated magnitude.
    \item Given $n$ and $P$ on the Gauss points, the $S$ field is then extracted by evaluating the normal `action' of $P$ on $n$ as
\begin{equation}\label{eq:S}
    S = \frac{1}{h}(Pn) \otimes n,
\end{equation}
where $h$ is the thickness of the interface, assumed here to be the resolution of the experimental data as discussed in Sec.~\ref{sec:expt}.

    We note that along an s-gb identified by the $|P|$ field, not all voxels are necessarily ``lighted up'' by a non-zero value of the $S$ field.
    \item With the $S$ field calculated on the Gauss points, the \textit{discrete} g.disclination density, $\mathit{\Pi}^{(d)}$ is defined at the center of each element. Recalling that $\mathit{\Pi} = \mbox{curl} \,S$ (and conducting the argument in 3-d for generality), for a pointwise orthornormal basis $(t_1, t_2, n)$, where $t_1, t_2$ are tangential to the interface, $\mathit{\Pi}^{(d)}$ is defined as

    \begin{equation*}
      \mathit{\Pi}^{(d)}  =  P \times n + (Pn) \otimes \mbox{curl}\, n
\end{equation*}
\begin{equation}\label{eq:Pxn}
         P \times n = (DW t_1) \otimes (t_1 \times  n) + (DW t_2) \otimes (t_2 \times  n)  + (DW n) \otimes (n \times  n).
\end{equation} 
\end{itemize}

As $\mathit{\Pi}^{(d)}$ is locally defined within an element, the $\mbox{curl}\,n$ term evaluates to zero, which also preserves a fundamental kinematical property of a smooth surface \cite[Appendix B]{ZAP}. From the representation of ($P \times n$) in (\ref{eq:Pxn}), it is evident that $\mathit{\Pi}^{(d)}$ is sensitive to only tangential gradients of `misorientation,' along the s-gb.

\begin{figure}[ht!]
\centering
    \includegraphics[scale=.70]{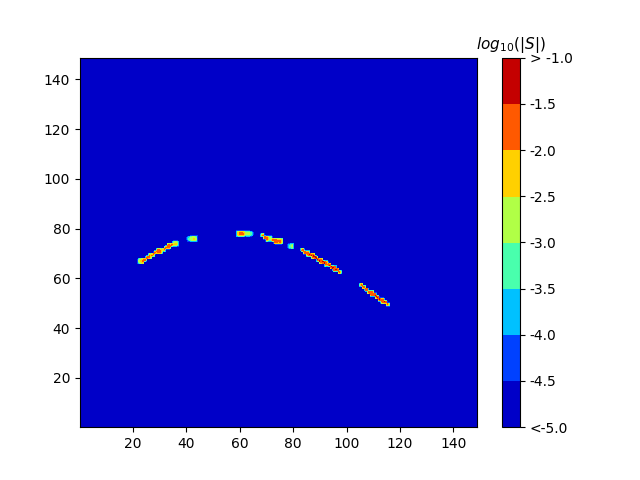}
    \caption{ $\log_{10} |S|$ in the domain for the isolated s-gb case. The background (domain apart from s-gb) has $S = 0$.}
    \label{fig:Norm of S field}%
\end{figure}

\begin{figure}[ht!]
    \centering
    \includegraphics[scale=.50]{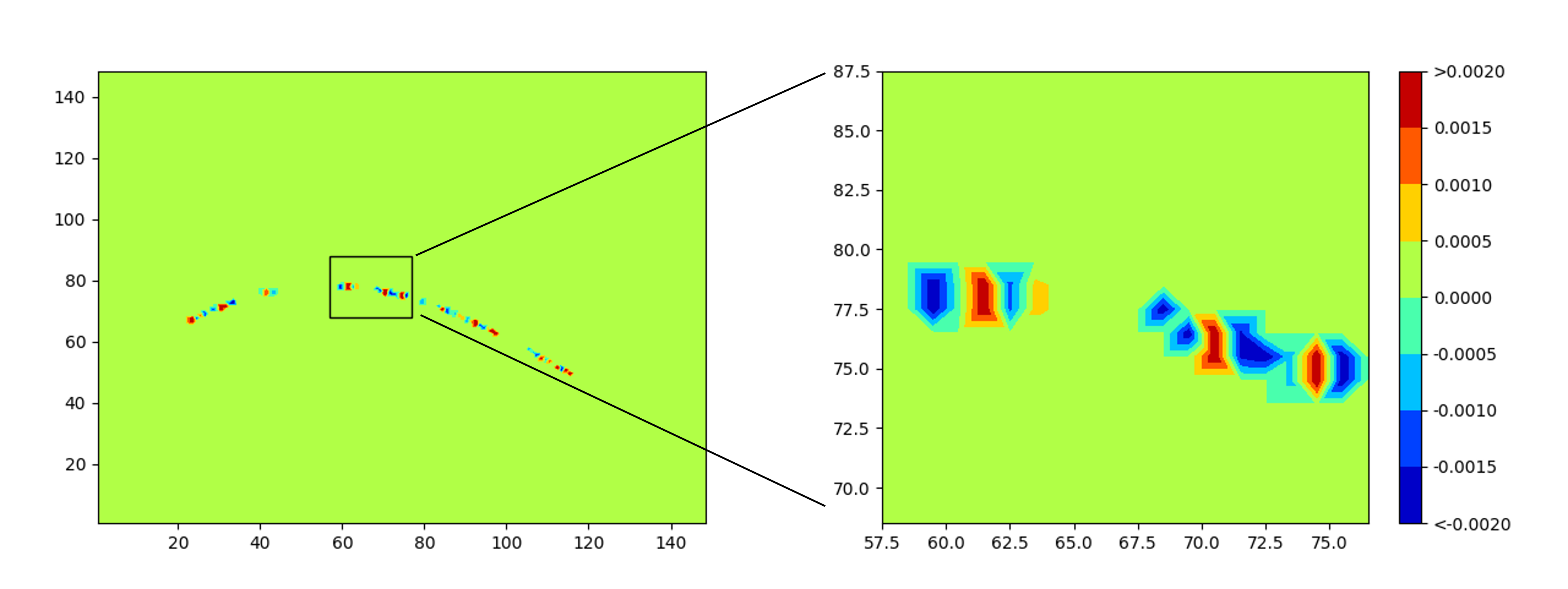}%
        \rput(-16,6){(a)}
        \rput(-8,6){(b)}
        
    \includegraphics[scale=.60]{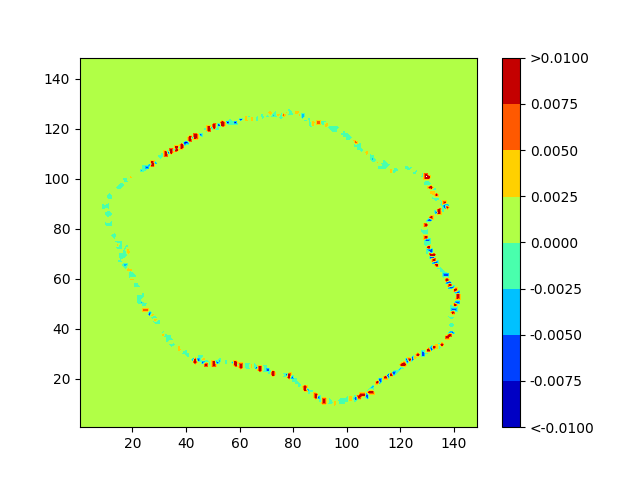}%
      \rput(-10,6){(c)}
    \caption{This figure illustrates the visualization of $\mathit{\Pi_{113}}^{(d)}$ for both (a) the s-gb and (c) the gb. To enhance the clarity of the dipole structure, a specific segment along the s-gb is zoomed in for detailed observation in (b).}%
    \label{fig:Dis_density}%
\end{figure}

Fig.~\ref{fig:Norm of S field} presents the spatial distribution of the $\log_{10} |S|$ field for the s-gb. In Fig.~\ref{fig:Dis_density}, we show plots representing a specific component of the $\mathit{\Pi}^{(d)}$ tensor. The figures show the s-gb as well as the single gb  that encloses the grain of interest. This distinct pattern, characterized by a g.disclination dipole structure, consistently emerges across all components of the $\mathit{\Pi}^{(d)}$ field at the same spatial locations. The spatial consistency in each component is evidence that the $\mathit{\Pi}^{(d)}$ field occurs as a tensor dipole along these interfaces. A g.disclination dipole is topologically equivalent to a dislocation, as shown in \cite{ZA, ZAP}.

The observed dipole structures are consistent with the expectation that the spacing of disclination dipoles/dislocations along a low-angle boundary (s-gb) is larger than in a high-angle boundaries (gb). While the defect structure along the gb is somewhat realistic, we recall that the $\mathit{\Pi}^{(d)}$ field is not used to calculating the elastic fields for reasons mentioned in Section~\ref{sec:modeling} related to accounting for finite rotations and deformations.

\subsection{Simulations and results}
\label{sec:simu_result}

\figref{fig:padding} shows the two domains and the features in them for which we perform simulations of the theory presented in Sec.~\ref{sec:g.disc_theory}. More specifically, \figref{fig:padding}(a) shows the s-gb identified from the experimental data and the $|P|$ threshold. This s-gb is defined based on orientation difference and confidence values for reconstruction. The s-gb has prescribed eigenwall fields processed from experimental data, as described above in Sec.~\ref{subsec:Cons_eig}. Elsewhere in the domain the eigenwall field is prescribed to be zero.
As this s-gb belongs to a polycrystal assembly, for the purpose of obtaining an estimate of the stress field and capture the effect of the network of grain boundaries, we included a larger elastic domain with the same elastic properties as used for the sample. Fig.~\ref{fig:padding}(b) shows the network of the grain boundaries along with the s-gb. The network of grain boundaries have non-zero interfacial dislocation density defined as discussed in Sec.~\ref{sec:modeling}. Elsewhere in the domain the dislocation density is set to zero.

We solve the stress equilibrium and elastic incompatibility equations, system (\ref{eq:GT_feq}), with zero-traction boundary conditions. Both isotropic and anisotropic simulations are carried out using a domain size of $150 \times 150$ voxels; each voxel is a square of edge length $1.8\ \mu m$. For either case, solutions for the elastic distortion are unique up to constant skew tensor, and this is assigned arbitrarily for the purpose of stress field evaluation.  If the distortion field is to be recovered, the known value of the experimental orientation field can be specified at one point of the domain. An important characteristic of our computational methodology, based on weak formulations of the governing equation (\ref{eq:GT_feq}) implemented by adapted algorithms within the finite element method, is that we do not have to calculate any derivatives of the $S$ field for the stress analysis, even though the calculations faithfully represent the stress field of the g.disclination field $\mathit{\Pi}$.

To simulate the anisotropic case, we need the components of the tensor of elastic moduli of each voxel with respect to the sample frame. For this we express the lattice basis frame in terms of the sample frame for every voxel. This linear combination between two orthogonal bases was directly obtained from the reconstructed experimental data. Thus,  if $(c_{i}), i = 1,2,3$ is the sample basis and $e^{A}_{i}$ is the lattice basis, and $A$ denotes a particular voxel,
\begin{eqnarray*}
\fl    \eqalign{& e^{(A)}_{i} = \alpha^{(A)}_{ij}c_j  \Rightarrow e^{(A)}_{i}\cdot c_{k} = \alpha^{(A)}_{ik}; \\
    &\mathbbm{C}^{(A)} = C^{(A)}_{ijkl} e^{(A)}_i\otimes e^{(A)}_j\otimes e^{(A)}_k\otimes e^{(A)}_l = \tilde{C}^{(A)}_{mnst} c_m\otimes c_n\otimes c_s\otimes c_t
    \Rightarrow \tilde{C}^{(A)}_{mnst} = C^{(A)}_{ijkl}\alpha^{(A)}_{im}\alpha^{(A)}_{jn}\alpha^{(A)}_{ks}\alpha^{(A)}_{lt}.\\}
\end{eqnarray*}

\begin{figure}[ht!]
    \centering
    \includegraphics[scale=.48]{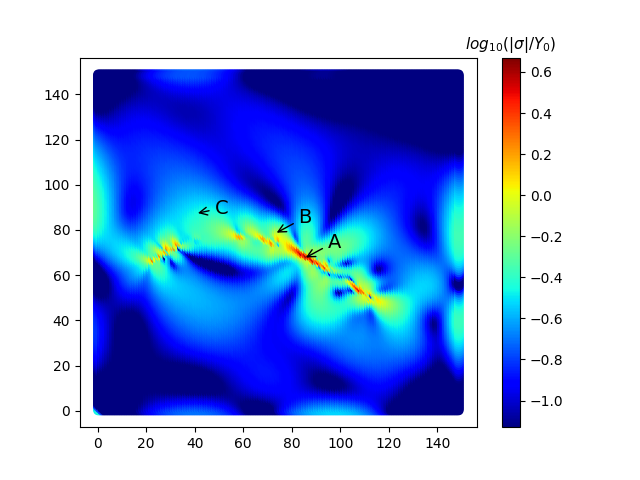}%
    \rput(-8,5){(a)}
    \qquad
    \includegraphics[scale=.48]{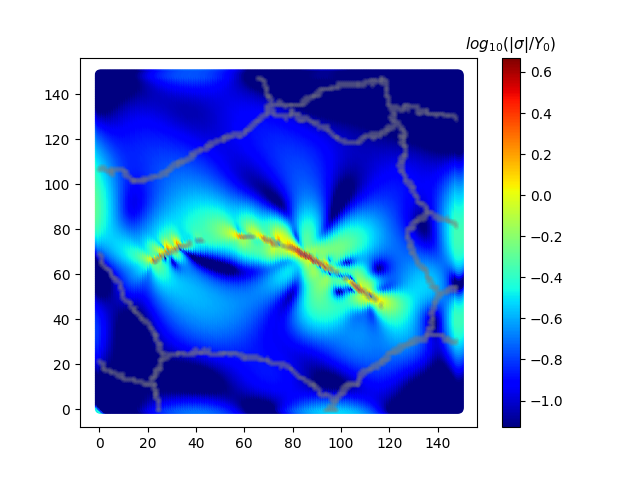}%
    \rput(-8,5){(b)}
    \caption{ Results for isotropic simulation is plotted for (a) the s-gb only and (b) the s-gb in the presence of large angle grain boundaries. The stress field is normalized by the yield stress so zero on the color scales correspond to a stress equal to the yield stress. Points $A$, $B$, $C$ are discussed in the text.}%
    \label{fig:Log_Stress_Norm}%
\end{figure}

\begin{figure}[ht!]
    \centering
    \includegraphics[scale=.48]{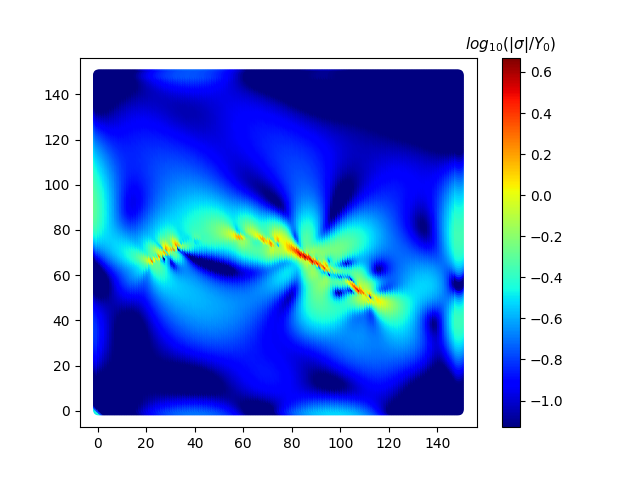}%
    \rput(-8,5){(a)}
    \qquad
    \includegraphics[scale=.48]{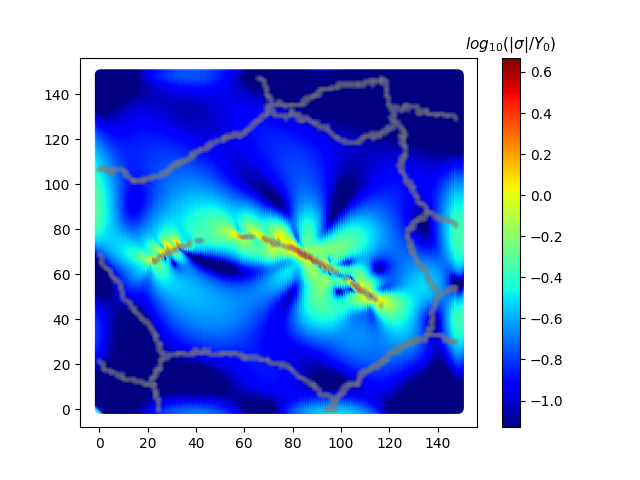}
    \rput(-8,5){(b)}
    \caption{ Results for anisotropic simulation is plotted for (a) the s-gb only and (b) the s-gb in the presence of large angle grain boundaries.}%
    \label{fig:Log_Stress_Norm_Aniso}%
\end{figure}

\begin{figure}[ht!]
\centering
    \includegraphics[scale=.70]{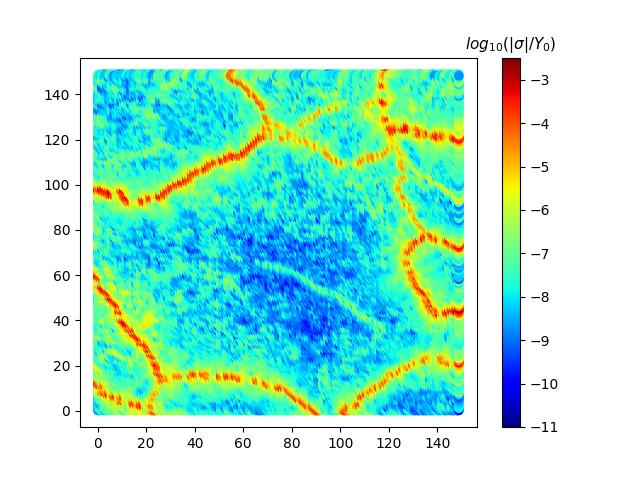}
    \caption{Result for isotropic simulation is plotted for just the network of grain boundaries calculated using skew-symmetric $U^e$. Note the different color scale used here.}
    \label{fig:zero_gb}%
\end{figure}

 
For both the isotropic (Fig.~\ref{fig:Log_Stress_Norm}) and anisotropic cases (Fig.~\ref{fig:Log_Stress_Norm_Aniso}), we find, as expected, that the stress field has large magnitude in the region where the $S$ field varies along the interface and especially at terminations. The stress field is almost the same in the `s-gb+gb'  and the `only-s-gb' cases due to the grain boundary network not producing significant stresses. Fig.~\ref{fig:zero_gb} shows the \mbox{log} of the norm of the stress field produced just by the network of grain boundaries; $Y_{0}$ denotes the yield stress for Zr, which is considered to be $150$ MPa \cite{2015AcMat..96..249G}. Despite the interfacial dislocation density of the gb network being highly inhomogeneous, vanishing stresses corresponding to it are calculated.

Reference points, $A$, $B$, and $C$, have been marked in Figure~\ref{fig:Log_Stress_Norm} to explain aspects of the stress field produced by the s-gb. The stress at point $A$, at a grain-size normalized distance of $0.004$ from the s-gb, has a yield normalized deviatoric (dev.) stress magnitude of $2.133$, sufficiently high to induce plasticity. Point $B$, at a distance of $0.042$, has a dev.~stress of $0.513$. Point $C$, is at a distance of $0.11$, and has a dev.~stress of $0.313$. Points $B$ and $C$ have stresses that can induce strong Bauschinger effects. The normalized total stress at each of these points is $4.173$ , $0.893$, and $0.5$, respectively, indicating the presence strong hydrostatic stresses with potential implications for void growth and closure. More specifically, at point $A$ the hydrostatic tension is 1.5 times the yield stress and can promote void growth. 

The predictions of such local stress inhomogeneities with strong implications for heterogeneous plasticity are contributions of our work, beyond the ambit of classical polycrystal plasticity.

The non-zero stress field away from the sub-grain boundary, in Fig.~\ref{fig:Log_Stress_Norm} and Fig.~\ref{fig:Log_Stress_Norm_Aniso}, is evidence of the fact that the field caused by a g.disclination extends well beyond its core. This observation can be further extended to infer that two g.disclinations can strongly interact even if they are macroscopically separated and, in fact, the force of this interaction is stronger than that of an equally separated pair of dislocations.
    
The stress fields for the isotropic and the anisotropic cases are similar in magnitude and spatial distribution, and are sufficient to induce plasticity. This observation holds for both the cases defined above. For the 2-d approximation used here (see Sec.~\ref{sec:2-d approx}) the elasticity tensor relates in-plane stress to the planar strain. For 2-d, $C_{ijkl}$, $i,j,k,l = {1,2}$ has 16 coefficients. Due to major and minor symmetries in the indices, we have 6 unique coefficients in the elasticity tensor. These coefficients are required to calculate the elastic fields. Even though the material is highly anisotropic, for the plane of interest used here (see Sec.~\ref{sec:2-d approx}) these coefficients are similar for both the cases. This yields similar elastic fields.

\section{Discussion}
An HEDM measured dataset has been analyzed to isolate s-gb features based on discontinuities (at the scale of observation) in the orientation field. These features have then been used to calculate associated elastic fields. Sources for the elastic fields arise from longitudinal variations of the rotation jump across the boundary, specified from the analysis of the experimental data.

Limitations in calculating the exact shape of the feature arise from the disparity in s-gb thicknesses and the resolution of the observations. Further,  noise in the measured orientation values will tend to generate computed stresses. For example in Fig.~\ref{fig:Norm of S field}, while longer length scale variations appear systematic, there are local (nearest neighbor) variations which may or may not be. Characterization of the results of various smoothed structures will be evaluated in future work.

Further limitations of translating available experimental data to input for elastic field calculations lies in the manual intervention required to determine an s-gb manifold (in this work, e.g., the calculation of the tangent along a best approximated s-gb curve as shown in Fig.~\ref{fig:sgb_prof}). More fully automated procedures can be developed to locate and characterize features throughout the measured volume of microstructure.

The calculations of defect elastic fields have been restricted to 2-d, small deformation, linear elasticity. A computational capability for 2-d, finite deformation calculations of g.disclination defect fields is available to us \cite{ZAP}. As well, finite strain, dislocation plasticity  codes in 2 and 3-d have also been developed by us \cite{Arora_acharya, Arora_zhang_acharya, Arora_acharya_2}. While theory and algorithms for 3-d, g.disclination based elastic field calculations have been developed by us \cite{ZAP}, a computational implementation for 3-d calculations needs to be developed. Finally, dislocation plasticity calculations in 3-d are computationally expensive, and more efficient strategies for parallel computation are needed to realize the potential of our combined experimental-theoretical advances.

\section{Conclusion}
This proof-of-principle calculation demonstrates a method that takes advantage of spatially resolved experimental measurements of unit cell orientation fields in three dimensions. The measurements include intra-granular orientation variations as well as rotations that naturally occur at grain boundaries. We calculate purely elastic stress fields of the defects that arise due to structural heterogeneities in orientation fields, without incorporating any plastic relaxation in this effort. Obviously, this can be improved. But these stresses represent precursor driving forces for localized, heterogeneous plasticity in and around sub-grain boundaries that perforce must exist in the material, and be mitigated/relaxed, since the orientation fields they correspond to persist for long time periods after real-time, step-wise tensile extension of the sample.

Future experimental analysis work will attempt to include the evaluation of inhomogeneous strain fields within grains \cite{Shen2020}. One example of a topologically interesting unit cell orientation discontinuity has been analyzed here. Future modeling work will extend the calculations to three dimensions and apply the methodology to cases where multiple orientation discontinuities occur within a single grain. A stretch goal is to use the full power of mesoscale field dislocation mechanics \cite{Arora_acharya, Arora_zhang_acharya, Arora_acharya_2} to model sample evolution from the S0 state to S2 of Fig.~\ref{fig:S0S2} and/or analogous evolution in other materials. Such exercises will help shed light on the prediction capabilities of our theoretical-computational models leading to their further refinement, as well as raise interesting questions for extending experimental measurements to more incremental sequences of states and to the use of complimentary techniques with higher spatial resolution in nanoscale grains in small samples \cite{Ulvestad2015}.

\section*{Acknowledgments}
This work was supported by the grant NSF OIA-DMR \#2021019. The authors declare that they have no known competing financial interests or personal relationships that could appear to have influenced the work reported in this paper.

\appendix
\section{Dislocation density of an infinitesimal rotation field cannot capture terminating interfaces}
\label{app:A}

We show here that interfacial dislocation density fields calculated from skew-symmetric elastic distortion fields as their curl cannot represent terminating interfaces (defined as formed from a rapid variation of the elastic rotation field over a small width, as for a gb or an s-gb).

Let the axial vector field of the skew symmetric tensor field $\omega^{(e)}$ be $\Phi$:
\begin{equation*}
 \omega^e = X \, \Phi; \qquad (\omega^e_{ij} = \epsilon_{ijk}\Phi_{k}). 
\end{equation*}
We will refer to the interfacial dislocation density field here as $\alpha^{(i)}$. It is closely related to the object $S:X$ discussed in Sec.~\ref{sec:g.disc_theory}; for an interface defined by a transition between two planar elastic distortion fields $U^1$ and $U^2$ in the direction $n$ over a width $h$
\begin{equation*}
        U_{ij,k} \approx \frac{U^1_{ij} - U^2_{ij}}{h} n_k; \qquad 
        \alpha^{(i)}_{il} = (\mbox{curl} U)_{il} = \epsilon_{lkj} U_{ij,k} \approx - \frac{(U^1 - U^2)_{ij} n_k}{h}  \epsilon_{jkl}.
\end{equation*}
When the elastic distortion field is skew,
\begin{equation*}
    \alpha^{(i)} = \mbox{curl}(\omega^e) = \mbox{curl}(X \Phi) =(\mbox{grad}\Phi)^T - (\mbox{div}\Phi) I; \qquad  (\alpha^{(i)}_{im} = \epsilon_{mjk} \omega^e_{ik,j}
 = \Phi_{m,i} - \delta_{im}\Phi_{n,n}),
\end{equation*}
where $I$ is the identity tensor.
Let $\Phi = \phi e_3$, where $e_3$ is the out-of-plane direction.
Consider now an arbitrary parametrization of the plane by the coordinates $\left(\xi^1 , \xi^2 \right)$ with natural basis denoted by $\left(a_i := \frac{\partial x}{\partial \xi^i}, i = 1,2 \right)$ and dual basis $\left(a^i := \frac{\partial \xi^i}{\partial x} \right)$. Then
\begin{eqnarray*}
    \alpha^{(i)} &= (\mbox{grad}(\phi e_3))^T - (\mbox{div}(\phi e_3))\mbox{I}\\
    &= \left[\frac{\partial(\phi e_3)}{\partial \xi^i} \otimes \frac{\partial \xi^i}{\partial x}\right]^T - \left[\mbox{tr}\left(\frac{\partial{(\phi e_3})}{\partial \xi^i}\otimes  \frac{\partial{\xi^j}}{\partial x}\right)\right]\mbox{I}\\
    &= \left[\frac{\partial \xi^i}{\partial x}\otimes \frac{\partial(\phi e_3)}{\partial \xi^i} \right] - \left[\frac{\partial{\phi}}{\partial \xi^i} e_3.  \frac{\partial{\xi^j}}{\partial x}\right]\mbox{I}
\end{eqnarray*}
and,  since both the primal and dual basis vectors are in-plane and perpendicular to the $e_3$ direction,
\begin{equation}\label{eq:alphai}
    \alpha^{(i)}_{i3} a^i \otimes e_3 = \alpha^{(i)} = \frac{\partial \phi}{\partial \xi^i} \left[a^i\otimes e_3 \right] \Longrightarrow \alpha^{(i)}_{i3} = \frac{\partial \phi}{\partial \xi^i},
\end{equation}
where $\alpha^{(i)}_{i3}$ are the non zero components of the interfacial dislocation density tensor on the (generally spatially varying) dual basis $a^i \otimes e_3, i = 1,2$.

By the equality of mixed partial derivatives, we then have
\begin{equation}\label{eq:mixed}
    \alpha_{13,2} =  \alpha_{23,1}.
\end{equation}
Thus, \textit{the components, w.r.t an arbitrary coordinate system, of the interfacial dislocation density tensor generated from a rotation field must satisfy the relation (\ref{eq:mixed}).}

The condition (\ref{eq:mixed}) holds in any coordinate system, including a rectangular Cartesian one. Geometric intuition for the condition is obtained by defining the vector field  $\rho := \hat{\alpha}_{13}\e_{1}  + \hat{\alpha}_{23}\e_{2}$, where the $\hat{\alpha}_{i3} = \alpha_{i3}$ for $(\xi^1, \xi^2)$ a rectangular Cartesian coordinate system with basis $(e_1, e_2)$. Then
\begin{equation}\label{eq:curlrho}
    \hat{\alpha}_{13,2} - \hat{\alpha}_{23,1} = 0 \Rightarrow \mbox{curl}\, \rho = 0 \Rightarrow \rho = \mbox{grad} \, \theta
\end{equation}
(we assume the region to be simply-connected) for some scalar field $\theta$, and (\ref{eq:alphai}) implies that $\theta = \phi$ (up to an irrelevant constant).

\begin{figure}[ht!]
    \centering
    \includegraphics[scale=.8
]{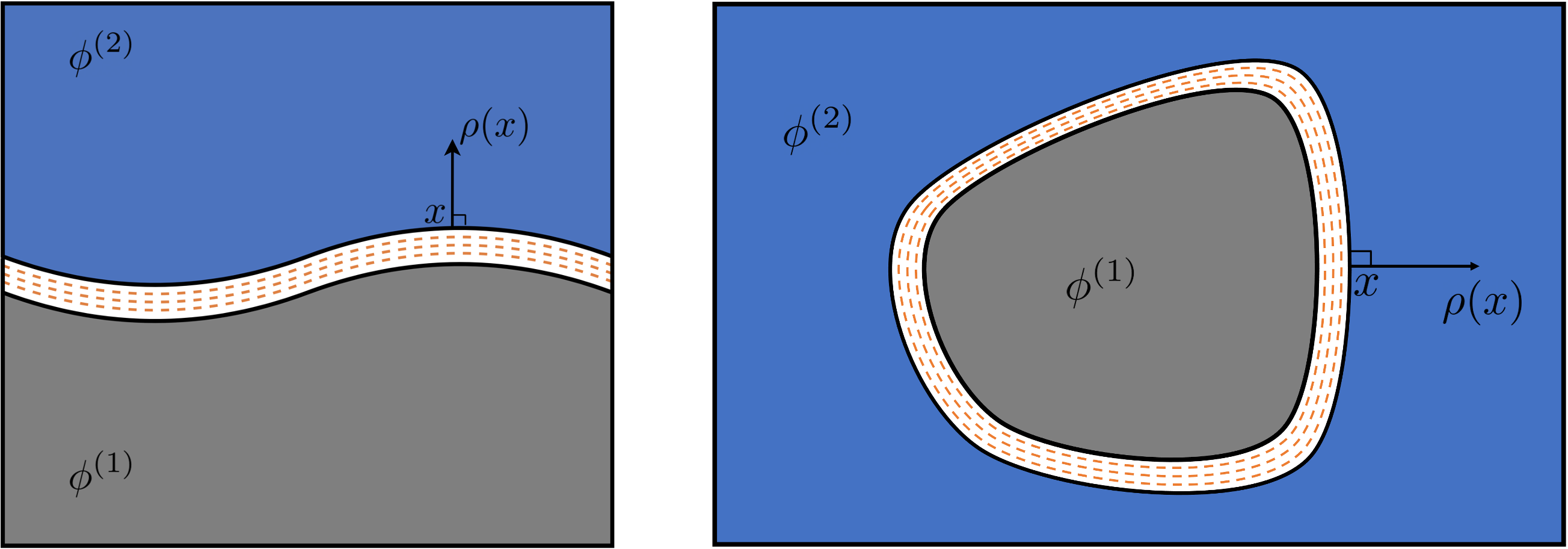}%
    \caption{A curved layer consisting of level sets of $\phi$ with the normal  representing the direction of $\rho$ at that point.}
    \label{fig:sch_edges_2}
\end{figure}

This geometrically means that the vector field $\rho$ points in the normal direction to level sets (contours) of the scalar field $\phi$, as shown in Fig.~\ref{fig:sch_edges_2}.

We now show that a layer with a specified interfacial dislocation density generated from a skew symmetric elastic distortion field cannot terminate, which is equivalent to the statement that if a layer interfacial dislocation density field terminates, it is not generated from a skew symmetric elastic distortion field (i.e. an infinitesimal rotation field).

\begin{figure}[ht!]
    \centering
    \includegraphics[scale=.5]{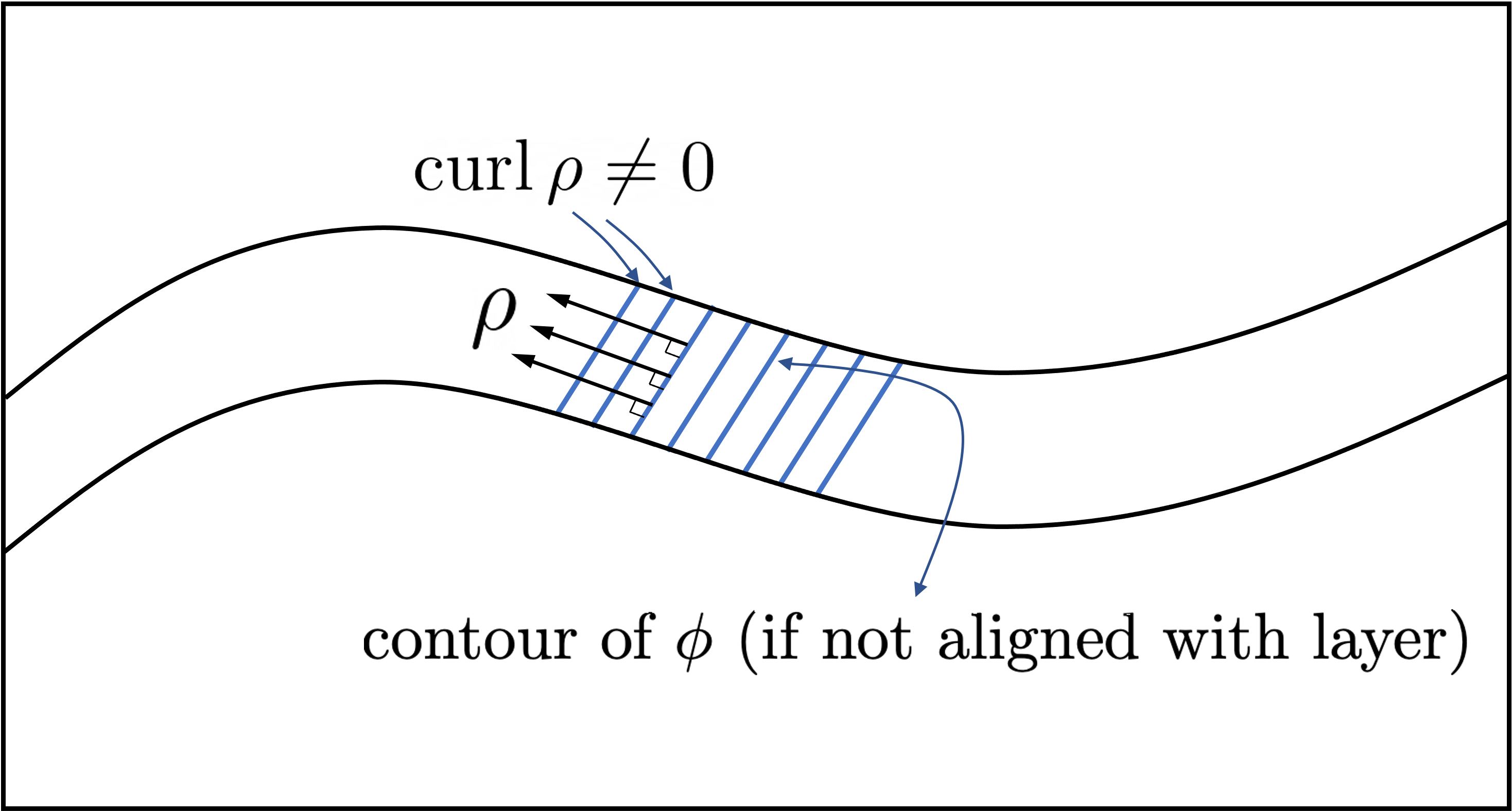}%
    \caption{Sketch for proving $\phi$ contours have to be parallel to layer by contradiction. $\rho$ has a component along the layer producing non-zero $\mbox{curl} \,\rho$.}
    \label{fig:appen_arg_1}
\end{figure}

Assume that an interfacial layer is generated from a skew-symmetric $\omega^e$. This means that the field $\rho = \mbox{grad} \,\phi$ is concentrated in the layer. But then the level contours of $\phi$ must be parallel to the layer, as otherwise the contours of $\phi$ would have to terminate at the lateral boundaries of the layer suggesting that $\mbox{curl} \, \rho$ would have to be non-vanishing at these boundaries (since there would be a gradient, in the direction parallel to the contours, of the $\rho$ field, which is normal to the contours), and this is not possible by (\ref{eq:curlrho}) - see Fig.~\ref{fig:appen_arg_1}. But, by the same logic, since the contours of $\phi$ are parallel to the (generally curved) interface, the interface itself cannot terminate as again, in that case, $\mbox{curl}\, \rho \neq 0$ at such a termination, and this is impossible. This proves the claim.

Thus, quantitatively speaking, $\mbox{curl}\, \alpha^{(i)T}$ may be considered as an indicator of layer termination and this field must vanish when the interfacial dislocation density field in question, $\alpha^{(i)}$, is generated from a skew symmetric field. This is consistent with the fact that in linear isotropic elasticity, the source of internal stress is given by the field $\left(\mbox{curl}\,\alpha^{(i)T} \right)_{sym}$ \cite{kroner1981continuum}.

A related characterization of the result is as follows: assume that a terminating boundary has a distribution of interfacial dislocation density on it in the general class indicated below in (\ref{eq:alphatilde}). Then it cannot satisfy the condition (\ref{eq:mixed}),  and hence it cannot originate from a skew-symmetric elastic distortion field. This is equivalent to the assertion that if an interfacial dislocation density originates from a skew-symmetric $U^e$ field then it cannot belong to the class (\ref{eq:alphatilde}). The argument is provided below.

\begin{figure}[ht!]
    \centering
    \includegraphics[scale=.5]{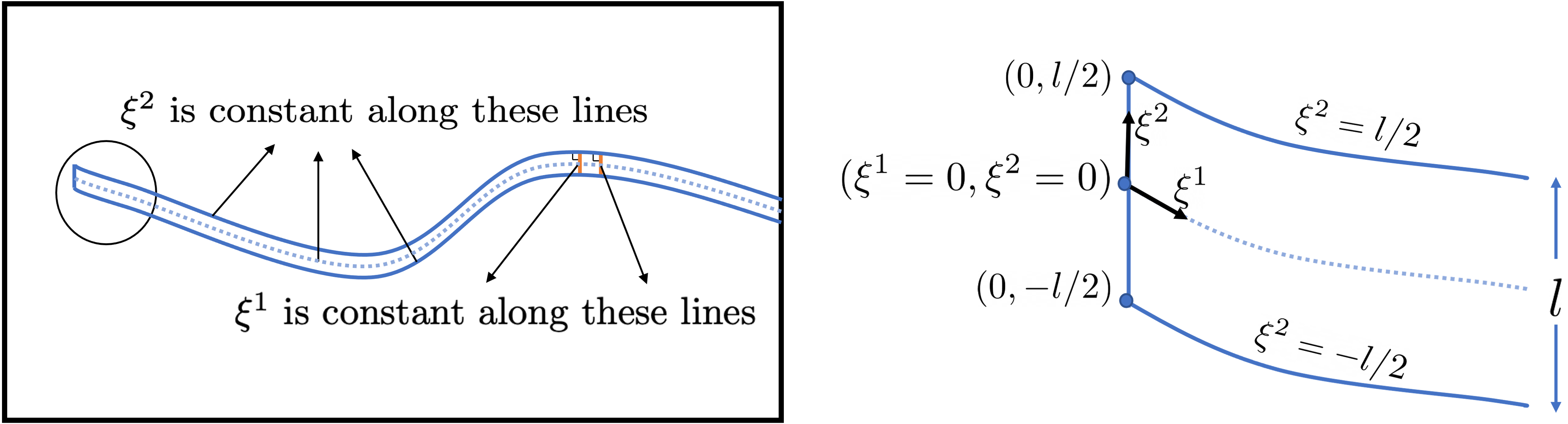}%
    \caption{Parametrization of a terminating layer by $\xi^1$ and $\xi^2$, and a zoomed in version at the termination.}
    \label{fig:appen_arg_2}
\end{figure}

Let $\tilde{\alpha} = \tilde{\alpha}_{13} \,a^1 \otimes e_3 + \tilde{\alpha}_{23} \, a^2 \otimes e_3 $ be a specified dislocation density on a terminating layer parametrized by curvilinear coordinates ($\xi^1, \xi^2 $) such that $\xi^2 $ is constant along the layer, and the components of $\tilde{\alpha}$ on the dual basis $a^i$ are given by
\begin{eqnarray}\label{eq:alphatilde}
    \tilde{\alpha}_{13}(\xi^1,\xi^2) &= A_1H(\xi^1)[H(\xi^2 - l/2) - H(\xi^2 + l/2) ] \nonumber\\
    \tilde{\alpha}_{23}(\xi^1,\xi^2) &= A_2H(\xi^1)[H(\xi^2 - l/2) - H(\xi^2 + l/2) ],
\end{eqnarray}
where $l$ is the thickness of the layer and $A_1$ and $A_2$ are constants, see Fig.~\ref{fig:appen_arg_2}.
$H(x - x_0)$ represent a Heaviside step function centered at $x_0$. Then,
\begin{eqnarray*}
    \label{eq:par_alpah}
    \tilde{\alpha}_{13,2}(\xi^1,\xi^2) &= A_1 H(\xi^1) [\delta(\xi^2 - l/2) - \delta(\xi^2 + l/2) ]\\
    \tilde{\alpha}_{23,1}(\xi^1,\xi^2) &= A_2 \delta(\xi^1) [H(\xi^2 - l/2) - H(\xi^2 + l/2) ]
\end{eqnarray*}
where $\delta(x - x_0)$ represents a Dirac delta function centered at $x_0$, and it is clear that $\tilde{\alpha}_{23,1}$ cannot equal $\tilde{\alpha}_{13,2}$, irrespective of the values for the constants $A_1$ and $A_2$ (except when both are zero) (this argument does not change in substance if $A_i$ are assumed to be functions of $\xi^1$).

\section*{References}
\bibliography{Singh.bib}
\bibliographystyle{unsrt}
\end{document}